\documentclass[10pt, a4paper, journal,twocolumn]{IEEEtran}

\IEEEoverridecommandlockouts

\usepackage{setspace}
\usepackage[pdftex, a4paper]{geometry}
\usepackage{graphicx}	
\usepackage{paralist}
\usepackage{pdfsync}
\usepackage{paralist}
\usepackage{tabularx,algorithm}	
\usepackage{color}
\usepackage{fancyhdr}
\usepackage{algpseudocode}
\usepackage{bm,bbm}
\usepackage{tikz}
\usepackage{paralist}
\usepackage{amsmath}
\usepackage{amsfonts}
\usepackage{dsfont}
\usepackage{mathrsfs}
\usepackage{amssymb}
\usepackage{bbm}
\usepackage{epsfig}
\usepackage[english]{babel}
\usepackage[all]{xy}

\newtheorem{theorem}{Theorem}
\newtheorem{definition}{Definition}

\newtheorem{lemma}{Lemma}
\newtheorem{remark}{Remark}
\newtheorem{corollary}{Corollary}
\newtheorem{proposition}{Proposition}
\newtheorem{example}{Example}
\newtheorem{assumption}{Assumption}

\def\diag{\mathop{\rm diag}}

\DeclareFontFamily{OT1}{pzc}{}
\DeclareFontShape{OT1}{pzc}{m}{it}{<-> s * [1.200] pzcmi7t}{}
\DeclareMathAlphabet{\mathpzc}{OT1}{pzc}{m}{it}

\newcommand*\mcapinn[2]{\vcenter{\hbox{$\mathsurround=0pt
  \ifx\displaystyle#1\textstyle\else#1\fi\bigcap$}}}

\newcommand*\mcupinn[2]{\vcenter{\hbox{$\mathsurround=0pt
  \ifx\displaystyle#1\textstyle\else#1\fi\bigcup$}}}

\title{ Localizability with Range-Difference Measurements }
\date{}

\author{Junfeng Wu, Biqiang Mu,
Xinlei Yi, Jieqiang Wei,
and Karl Henrik Johansson
\thanks{J. Wu is with the School of Data Science, the Chinese University of Hong Kong, Shenzhen, and the College of Control Science and Engineering and the State Key Laboratory of Industrial Control Technology, Zhejiang University, P. R. China. {Email: junfengwu@cuhk.edu.cn}. B. Mu is with the Key Laboratory of Systems and Control, Institute of Systems Science, Academy of Mathematics and Systems Science, Chinese Academy of Sciences, Beijing 100190, China.
{Email: bqmu@amss.ac.cn}. X. Yi and K. H. Johansson are with the Division of Decision and Control Systems, School of Electrical Engineering and Computer Science,
 KTH Royal Institute of Technology.
 {Emails: \{xinleiy, kallej\}@kth.se}. J. Wei is with Ericsson, Torshamnsgatan 21, 16440, Stockholm, Sweden.
{Email: jieqiang.wei@gmail.com}.}

}

\begin{document}
\maketitle

\begin{abstract}
The physical position is crucial  in location-aware services or protocols based on geographic information, where localization is performed given a set of sensor measurements for acquiring the position of an object with respect to
a certain coordinate system. {In this paper, we revisit the long-standing localization methods for locating a radiating source
from range-difference measurements, or equivalently, time-difference-of-arrival  measurements from the perspective of least squares (LS).
In particular, we focus on the spherical LS error model, where the error function is defined as the difference between the squared true
distance from a signal receiver (sensor) to the
source and its squared measured value, and the resulting spherical LS estimation problem.}
This problem has been known to be challenging due to the non-convex nature of the hyperbolic measurement model.
First of all,  we prove that the existence of least-square solutions is universal and that solutions are bounded under some assumption on the geometry of the sensor placement. Then a necessary and sufficient
condition is presented for the solution characterization based on the method of Lagrange
multipliers.  Next, we derive a characterization for the uniqueness of the solutions incorporating a second-order optimality condition. The solution structures for some special cases are also established, contributing to insights on the effects of the Lagrangian multipliers on global solutions.
These findings  establish a comprehensive understanding of
the localizability with range-difference measurements, which are also illustrated with numerical examples.
\end{abstract}

\section{Introduction}

{
Location-based services and protocols~\cite{gentile2012geolocation} can be seen in a
large number of applications, ranging from wireless communication~\cite{mao2007wireless,win2011network,huang13twc}, internet-of-things~\cite{vermesan2011internet}, transportation~\cite{papadimitratos2009vehicular} to advertising or social networks~\cite{constandache2010did}. It is worthwhile to note that, amid the proliferation of localization-based applications in mobile networks, we have seen constantly surging interests in localization. This trend has been boosting a batch of researches on
distributed localization~\cite{langendoen2003distributed,priyantha2003anchor,khan2009distributed,Canclini15IASLP},
indoor localization~\cite{youssef2005horus,chintalapudi2010indoor,rai2012zee}, simultaneous localization and mapping~\cite{dissanayake2001solution,durrant2006simultaneous}, to name a few.
Localization problems focus on acquiring the position of an object in a certain coordinate system based on measurement data from a set of sensors.
The methods for localization vary, among which localization using field geometry about objects' relative placement (also known as range-based localization
methods)~\cite{chandrasekhar2006localization,dil2006range} are practically prevalent since the geometric
information of an object can be easily measured by sensors.
{
The localization methods utilizing time difference of arrival (TDoA)~\cite{musicki2010mobile, salameh2010cooperative, karbasi2012robust, liu2007survey,tiemann2017scalable, bottigliero2021low,martalo2021improved,raza2019dataset,sidorenko2020error}
are important examples of the lateration approach for passive localization. It performs with high
accuracy since it does not rely on time synchronization between the source and receivers.}}

{
In this paper, we consider the problem of  locating a radiating
source
from TDoA measurements.
In an ideal uniform medium,  radiation
 travels through the medium at a known velocity, and therefore
the measurement of range differences from the source is accessible from the TDoA measurement.
{Source localization from range-difference measurements has been extensively studied by means of least squares methods. In particular,
finding the least squares solution with respect to the spherical error criterion~\cite{huang2001real} can be formulated as an optimization problem with a quadratic objective function and some constraints, also known as a constrained least square (CLS) range-difference based localization problem.
The problem is nonconvex as the Hessian matrix of a quadratic term of one constraint is not positive semi-definite. It results in difficulty in finding a global solution.

Finding solutions for the CLS problem and its variants has been an active research topic in the area of localization.
A direct method
is discarding the quadratic constraints~\cite{li2004least,Stoica2006}, which gives rise to an unconstrained least squares (ULS) problem.
An extension of
 of the ULS
method  is to explore the constraint to
reduce the error of the ULS solution for generating the final
estimate~\cite{chan1994simple,sun18tsp}.
The spherical-interpolation method is a commonly used approach
proposed that solves the CLS problem approximately~\cite{Smith1987IJOE,smith1987closed,Stoica2006}, by which optimization is
done by alternating restricted minimization over two disjoint subsets of the variables.
The subspace minimization method~\cite{smith1987closed,Stoica2006}
is
another way to solve the CLS problem approximately. In this method, orthogonal projection is used to eliminate the challenging quadratic constraint.
As a revisit to the methods reviewed above, it is pointed by~\cite{Stoica2006} that
the spherical-interpolation method and the subspace minimization method are identical to
the unconstrained LS method in the sense that these methods generate identical solutions.
The reference~\cite{Schau1987ITASSP} presents a closed-form localization technique, termed the spherical-intersection method, which has similar formulation to the spherical-interpolation one. What is worth noticing is that
the spherical-intersection method gives consideration to quadratic constraint but
fails in obtaining a global minimizer too as it by its nature is alternating optimization. Some solvers are iterative methods, such as~\cite{Foy76TAES,Torrieri84TAES}. At each step a location estimate is improved by
solving a local LS estimation problem using the Taylor-series method. However, proper
initialization is needed to avoid false local minima~\cite{sun18tsp}.
A different category of solvers for the CLS problem is
based on the Lagrange multiplier technique.
In~\cite{Beck2008,sun18tsp}, numerical global solution searching algorithms are developed in virtute of necessary conditions for the global optimality of the CLS problem, as a consequence of the  the analysis on generalized trust region problems offered in~\cite{More93OMS}. The solution finding amounts to finding roots of a $4$th order polynomial in the two-dimensional or a $6$th
order in the three-dimensional localizations. Such extensions are
possible because the CLS problem is closed to
a generalized trust region problem in form. To attack the global solution, the methods in~\cite{Beck2008,sun18tsp} need an exhaustive search for all the suspected, which is numerically less efficient. 
}}

{ There are a few key questions remaining to be answered for this range-difference based localizations: (i) How are the solutions to the CLS problem related to the localization measurements?
	 (ii) When does the CLS problem  for range-difference based localizations  have a global/unique solution? (iii) How  can the global/uniq
     
        ue  solution be characterized?}
	 In this paper, we establish a series of results attempting to address these questions for range-difference based localization problems utilizing the persistence of excitation on the localization measurements as well as an extended and structured  Karush-Kuhn-Tucker (KKT) analysis. The main contributions of our paper are threefold as follows.
{
\begin{enumerate}

\item [$(i)$.] We show that in the absence of measurement noises, the persistence of excitation on the localization measurements describes whether or not the coordinate of a radiating source can be exactly recovered from
TDoA measurements, which depends on the rank of  the Jacobi matrix of the model, and that for a practical CLS problem where measurement noise is taken into account
the CLS solutions always exist and under some assumption on the Jacobi matrix of the measurement model the solution set is bounded.
In principle, the condition suggests that a larger-sized sensor array inclines to lead to a bounded CLS solution.

\item [$(ii)$.]
We develop a necessary and sufficient condition for a global CLS solution in terms of a group of KKT conditions.
Moreover, we also establish a characterization for the uniqueness of an optimal solution.  The uniqueness result is attached to a stricter condition on the curvature of a Lagrangian function.

\item [$(iii)$.]The theoretical results are developed in this paper inspire us to
consider some special cases, and uncover insightful findings on the position of the Lagrangian multiplier affecting how the global minimizers can be solved in these cases. They pave a way for us to compute a global minimizer in a relative easy manner. In this part, we also draw remarks on the multiplicity nature
of CLS solutions.

\end{enumerate}
}
Simulation examples are carried out to illustrate our theory. The derivation of our results are partly inspired
by~\cite{More93OMS}. In particular, the reference~\cite{More93OMS}
gives characterization of
the global minimizer of the generalized trust region problem and its uniqueness
in terms of the Lagrange multiplier. However, the additional positivity constraint, see~\eqref{eqn:opt-problem-LS-trans-constr2}, leads to challenging difficulty in analysis. The global optimal solution has been investigated in~\cite{Beck2008}, and a
sufficient condition and a necessary one are offered
separatively. The two conditions do not coincide in general and
contain conservativeness, which can be seen from a concrete example in that work.
In contrast, our results close the gap.
In addition, we develop a condition for the solution uniqueness. To the best of our knowledge, this is the first time that
a characterization of the uniqueness is given in the literature.

The remainder of the paper is organized as follows.
In Section~\ref{section:problem-statement}, we introduce range-difference localization and how it is formulated into a CLS localization problem, as well as the problems of interest in the paper.
In
Section~\ref{section:characterization}, we develop characterizations of solutions to the CLS localization. Specifically, the development mainly includes a feasibility condition for the problem, a characterization of a global solution and a uniqueness characterization.
We also give comparison between our results and the main related ones in literature.
In Section~\ref{section:solution_algorithm}, we establish a few findings on the structural properties of global solutions in some special cases.  We also give some numerical examples to illustrate our theory. Finally, some concluding remarks
are drawn in Section~\ref{section:conclusions}. Most of the proofs of the main results can be found in Appendices.

\noindent\textbf{Notations}.~We use $\sigma_{\rm min}(X)$ and $\sigma_{\rm max}(X)$ to denote the eigenvalues of a square real matrix $X$, which have the smallest and largest magnitude, respectively.
For $x,y\in\mathbb{R}$, $x\vee y$ and  $x\wedge y$ stand for
the maximum and minimum of $x$ and $y$, respectively.
For a real-valued function $h(x): \mathbb R^{n}\rightarrow \mathbb R$,
we use $\nabla h$ to denote the gradient of $h$.
For a set $\mathcal A$, we use $\overline{\mathcal A}$ to denote the closure of
$\mathcal A$. Let $\mathcal A$ be a metric space, which is a
set equipped with a metric $d$.
Suppose that $ \emptyset\not =\mathcal B \subset  \mathcal A$ and
$x\in\mathcal A$. The distance of $x$ from $\mathcal A$ is defined as
$d(x;\mathcal A)=\inf\{d(x,y):y\in \mathcal A\}.$

\section{Problem Statement}\label{section:problem-statement}

{
\subsection{Range-difference Based Localization}
We consider a radiating source and an array consisting of $m+1$ sensors that collect signals emitting from the source.
We
denote the coordinate, with respect to an Euclidean coordinate system, of the source by $x\in\mathbb R^n$
and denote
the coordinate of
sensor $i$ by $a_i\in\mathbb R^n$. In particular we assume $a_0=0$ for sensor $0$, i.e.,
sensor $0$ is set to be at the origin of the coordinate system. We use $d_i$ to denote the
range-difference measurement from sensor $i$, for $i=1,\ldots,m$, to sensor $0$.
 We adopt the additive measurement error model, in which  the measurements of the range differences are modeled as
$d_i$, $a_i$ and $x$ can be given as
\begin{equation}\label{eqn:di}
d_i=\|a_i-x\|-\|x\|+r_i,
\end{equation}
where $\|\cdot\|$
denotes the $\ell_2$ norm of a vector and $r_i$ captures the measurement noise contained in $d_i$, which is also called ``equation error''~\cite{ljung1983theory}. Therefore, a natural localization problem arises, on identifying the unknown source $x$ from the measurement data $\{d_i\}$ and the sensor localizations $\{a_i\}$. { We postulate that the additive measurement errors have
mean zero and are independent of the range difference observation and the source location, as routinely assumed in the literature, such as~\cite{huang2001real}.}

\begin{figure}
\centering
\begin{tikzpicture}
\draw[black] (-1,0) -- (5,0);
\draw[black] (0,0) circle (2.5pt) node[anchor=north] {$a_0=0$};
\draw[black] (2,0) circle (2.5pt) node[anchor=north] {$a_1$};
\draw[black] (4,0) circle (2.5pt) node[anchor=north] {$a_2$};
\filldraw[black] (-1,3) circle (2.5pt) node[anchor=south] {$x$};
\draw[black] (-1,3) -- (0,0);
\draw[black] (-1,3) -- (2,0);
\draw[black] (-1,3) -- (4,0) ;
\draw[] (0,0) arc (288.4349:340:3.1623);
\draw [black] (2,0) -- (2.2,0.2);
\draw [black] (1.24,0.76) -- (1.44,0.96);
\draw[ <->] (2.1,0.1) -- (1.34,0.86) node[midway,above] {$d_1$};
\draw[ black] (1.73,1.36) -- (1.88,1.61);
\draw[ black] (4,0) -- (4.15,0.25);
\draw[ <->] (4.075, 0.125) -- (1.805,1.485)  node[midway,above] {$d_2$};
\end{tikzpicture}
\caption{Illustration of range-difference measurements in an array consisting of three sensors (i.e., $m=2$). The array is a line one. The solid dot ``$\bullet$'' represents the radiating source and the hollow dots ``$\circ$'' represent the sensors.}
\label{fig:figure_TDoA}
\end{figure}
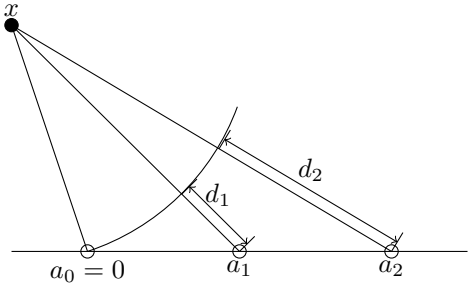

The TDoA techniques, as means of passive localization, has a large number of applications in different positioning systems~\cite{leonard2012directed,sundar2018tdoa,le2019uncovering,cobos2020frequency}. For instance, they have been widely used for sound source localization,
where the goal is to estimate the coordinates of
a sound source using acoustic signals received by an array
of microphones. The microphones are mounted at fixed and known positions $a_i$, with one of them selected as the reference node. The sound source locates
at an unknown position $x$.
The TDOA measurement $d_i$ is actually time delay that the sound takes to get received by the $i$th and the reference nodes (microphones), multiplied by the
propagation speed of sound in appropriate medium, see Figure~\ref{fig:figure_TDoA}. Then a corresponding hyperbolic surface can be derived from a TDoA measurement, for which all points on the surface will have the same distance difference from the said pair of nodes.


\subsection{Persistence of Excitation of the Spherical LS Error
Model}

{
The spherical LS model built on the distance error from the hypothesized source location to every sensor, a transformed model of~\eqref{eqn:di}, has a simpler expression on the localization $x$ than \eqref{eqn:di}.
Consider the distance from the source to the $i$th sensor. The measured value from the
TDoA measurement model~\eqref{eqn:di} can be computed as
$$d_{ix}=d_i+\|x\|$$
and the true value is $\|a_i-x\|$. The spherical LS error function is defined as the difference of the squares of $d_{ix}$ and $\|a_i-x\|$:
$$
e_i=\frac{1}{2}d_{ix}^2-\frac{1}{2}\|a_i-x\|^2.
$$
By~\eqref{eqn:di}, we obtain the spherical LS model
\begin{equation}\label{eqn:di-squared}
e_i=d_i\|x\|+a_i^\top x-b_i, \hbox{~for ~}i=1,\ldots,m,
\end{equation}
where $b_i=\frac{1}{2}(\|a_i\|^2-d_i^2)$. By~\eqref{eqn:di} and~\eqref{eqn:di-squared}, we have
$e_i=\|a_i-x\|r_i+\frac{r_i^2}{2}$.

Suppose that the measurement errors were absent (i.e., $r_i\equiv 0$).
Then model~\eqref{eqn:di} defines a branch of a hyperbola, of which $a_i$ and $a_0$ are the two foci and $d_i$ is the focal distance. 
For a practical TDoA problem, the true position of the source must exist, which means that the $m$ equations have at least one intersection. 
Meanwhile, the spherical LS error $e_i$ of the spherical LS  model  \eqref{eqn:di-squared} turns to be zero.
Here one basic concept {\it persistence of excitation} of a given set of
locations $\{a_i:i=0,\cdots,m\}$ at a source location $x\in\mathbb R^n$
lies in, whether or not the exact $x$ can be uniquely recovered from $\{a_i:i=0,\cdots,m\}$ in some sense.
When the answer is affirmative, we call the measurement locations $\{a_i:i=0,\cdots,m\}$ are persistently exciting (PE) at $x$.
The concept of persistence of excitation is highly related to 
localizability of network localization systems and rigidity of graphs as they are all about whether the location(s) of 
node(s) can be uniquely determined using edge information. 

Since the global persistence of excitation  is difficult to check for nonlinear models \eqref{eqn:di} and \eqref{eqn:di-squared},
here we instead introduce a little weaker condition (the local persistence of excitation conception) of input signals in system identification \cite{Ljung94Auto} and  derive a verifiable condition for the measurement locations $\{a_i:i=0,\cdots,m\}$ to a source location.
\begin{definition}\cite[
	Definition 3]{Ljung94Auto}
Consider the spherical LS model~\eqref{eqn:di-squared} in the absence of noises, i.e., $r_i\equiv 0$. The  set of measurement locations $\mathcal A:=\{a_i:i=0,\cdots,m\}$ is said to be locally PE with respect to~\eqref{eqn:di-squared} at a source location $x\in\mathbb R^n\setminus \mathcal A$ if there holds that
$d_i\|\tilde{x}\|+a_i^\top \tilde{x}-b_i=0, i=1,\ldots,m$
 for any $\tilde{x}$ in a sufficiently small neighborhood of $x$,
 then $\tilde x= x$.
\end{definition}
This local PE is a prerequisite for developing numerical algorithms to estimate $x$, otherwise intrinsic ambiguity persists: there are at least two different possible source locations  that correspond to measurements $d_i$ even in the absence of noises.
Actually, the local PE issue aims to find the condition on the measurment locations under which the solution to the spherical LS model (\ref{eqn:di-squared}) is unique in a small neighborhood of the true position of the source.

For models \eqref{eqn:di} and \eqref{eqn:di-squared}, in the absence of measurement noises, the measurement location set $\{a_i,i=0,\cdots,m\}$ is locally PE if  their Jacobian matrices at the true location
are of column full rank  \cite{Ljung}.
The Jacobian matrix $J_0(x)$ of~\eqref{eqn:di} is
\begin{align}\label{eqn:J0x}
J_0(x)&= -\left[\begin{array}{cc}
\frac{x^\top}{\|x\|}+\frac{(a_1-x)^\top}{\|a_1-x\|}\\
\vdots\\
\frac{x^\top}{\|x\|}+\frac{(a_m-x)^\top}{\|a_m-x\|}
\end{array}
\right],
\end{align}
while  the Jacobian matrix $J(x)$ of \eqref{eqn:di-squared} is:
\begingroup
\allowdisplaybreaks
\begin{align}
J(x)&=
\begin{bmatrix}
w_1\\
\vdots\\
w_m
\end{bmatrix}
\frac{x^T}{\|x\|}
+
\begin{bmatrix}
a_1^T\\
\vdots\\
a_m^T
\end{bmatrix}\label{eqn:Jx}  \\ &=\frac{1}{\|x\|}\left[\begin{array}{cc}
\|a_1-x\|x^\top+(a_1-x)^\top\|x\|\\
\vdots\\
\|a_m-x\|x^\top+(a_m-x)^\top\|x\|
\end{array}
\right]\notag \\
&= -\left[\begin{array}{ccc}
\|a_1-x\| & &\\
& \ddots &\\
&  & \|a_m-x\|
\end{array}\right]J_0(x)\notag
\end{align}
\endgroup
where $w_j=\|a_j-x\|-\|x\|$ for $j=1,\cdots,m$ are applied in the second equality.
This means that the model \eqref{eqn:di} is equivalent to the model \eqref{eqn:di-squared} in the absence of measurement noises when $x\neq a_i$ for all $i=1,\cdots,m$. The measurement locations $\{a_i,i=1,\cdots,m\}$ are locally PE if they satisfy special geometric relations, which is given in the following proposition.
{
\begin{proposition}
	Suppose that $r_i\equiv 0$.
The  measurement location set $\{a_i:i=0,\cdots,m\}$ is locally PE with respect to~\eqref{eqn:di-squared} at a generic source location $x\in\mathbb R^n$
 if
\begin{enumerate}
    \item for $n=2$, all of measurement locations $\{a_i:i=0,\cdots,m\}$ are not collinear;
    
    \item for $n=3$, $\{a_i:i=0,\cdots,m\}$ are not coplanar.
\end{enumerate}
\end{proposition}
\noindent \textbf{Proof.~}
By the inverse function theorem~\cite{Rudin76Math_analysis},  $J(x)$ being nonsingular implies that 
there exists  an
open neighborhood $\mathcal U$ around $x$ there is a diffeomorphism mapping $\mathcal U$ to
 $e(\mathcal U)$, which further implies local PE.
If  $\{a_i:i=1,\cdots,m\}$ is not locally PE with respect to~\eqref{eqn:di-squared} $x\in\mathbb R^n$, we have that
$J(x)$ (because $x\in\mathbb R^{n}/\ \mathcal A$ so does $J_0(x)$) does not have full column rank, that is, for any 
$i_1,\ldots,i_n\in\{1,\ldots,m\}$, there exist real numbers
$\beta_1,\ldots,\beta_n$ that are not all zero, such that
$$
\sum_{l=1}^n \beta_l(\frac{x}{\|x\|}+\frac{a_{i_l}-x}{\|
a_{i_l}-x\|})=0. 
$$
This is further equivalent to the condition that the rank of 
$$
[
\frac{x}{\|x\|}+\frac{a_{i_1}-x}{\|
a_{i_1}-x\|},\ldots,
\frac{x}{\|x\|}+\frac{a_{i_n}-x}{\|
a_{i_n}-x\|}]
$$
is strictly less than $n$, or
its determinant is zero. Let $\mathcal V$ be the set of $x$ resulting in vanishing determinants of all so defined matrices  associated with any $n$ location selections from $\{a_i:i=1,\ldots,m\}$. 
Since all of measurement locations $\{a_i:i=0,\cdots,m\}$ are not collinear for $n=2$ or not coplanar for $n=3$, there exists at least one collection, termed $i_1',\ldots, i_n'$, implying that vectors $a_{i_1'},\ldots, a_{i_n'}$ are linearly independent.  Therefore the set of $x$,  leading to that the determinant of 
$$
[
\frac{x}{\|x\|}+\frac{a_{i_1'}-x}{\|
a_{i_1'}-x\|},\ldots,
\frac{x}{\|x\|}+\frac{a_{i_n'}-x}{\|
a_{i_n'}-x\|}]
$$
vanishes, is proper, which further ensures $\mathcal V$ to be proper. Therefore the location set  $\{a_i:i=0,\cdots,m\}$ is locally PE with respect to~\eqref{eqn:di-squared} at a generic $x\in\mathbb R^n$, which completes the proof. \hfill $\blacksquare$
}}}



\subsection{Practical CLS Solution from the Spherical LS Error Model}
A fundamental approach to estimate $x$
 is to minimize the summed spherical least squares errors $e_i$'s, i.e.,
\begin{equation}\label{eqn:opt-problem-LS}
\hbox{(\textbf{LS}):}~~~~~~~\mathop{\rm minimize}_{x\in\mathbb R^n} \sum_{i=1}^{m} (d_i\|x\|+a_i^\top x-b_i)^2,
\end{equation}
The LS problem is nonconvex. Letting $y=[\|x\|,x^\top]^\top $, we transform the LS problem equivalently into
the following constrained least squares (CLS) problem
\begin{subequations}\label{eqn:opt-problem-LS-trans}
\begin{align}
\mathop{\rm minimize~}\limits_{y\in\mathbb R^{n+1}} & f(y)\label{eqn:opt-problem-LS-trans-obj}\hspace{30mm}\\
\hbox{\textbf{(CLS)}:}~~~~~~~~~\mathop{\rm subject~to~} & g(y)=0,\label{eqn:opt-problem-LS-trans-constr1}\\
& [y]_1\geq 0.\label{eqn:opt-problem-LS-trans-constr2}
\end{align}
\end{subequations}
In~this problem, $f(y):=\|Ay-b\|^2$, where
\begin{equation}\label{eqn:def-A}
A=\left[
\begin{array}{ccc}
d_1 & a_1^\top  \\
\vdots      & \vdots\\
d_m & a_m^\top
\end{array}
\right],
\end{equation}
$b=[b_1,\ldots,b_m]^\top$, $g(y):=y^\top D y$ with
\begin{equation}\label{eqn:def-D}
D=\mathop{\rm diag}(1,-I_{n}),
\end{equation}
where $I_n$ denotes an identity matrix of size $n$ and
${\rm diag}(\cdot,\ldots,\cdot)$ denotes a diagonal matrix with the arguments
in brackets on the main diagonal, and the notation
$[x]_i$, $i\leq p$, is used to denote the $i$th element of $x$ for any given
$x\in\mathbb R^p$. The interpretation of each equation in the CLS problem is
as follows: the least square criterion is given by~\eqref{eqn:opt-problem-LS-trans-constr1} and~\eqref{eqn:opt-problem-LS-trans-constr2} describe
the geometric constraints $\|x\|^2=[x]_1^2+\cdots+[x]_n^2$ and
$\|x\|\geq 0$. An optimal value $f^*$ of the CLS problem is defined as
\begin{equation}\label{eqn:f-optimal-definition}
f^*_{\rm CLS}=\inf\{f(y):[y]_1\geq 0, g(y)=0,y\in\mathbb R^{n+1}\}.
\end{equation}

\subsection{Problems of Interest}

To find the least squares solution with respect to the spherical error
       criterion~\cite{huang2001real}, it                                       resorts to solving the CLS problem~\eqref{eqn:opt-problem-LS-trans}, which was firstly studied in this reorganization idea in~\cite{Smith1987IJOE,smith1987closed,friedlander1987passive} and was investigated
widely in later work, such as~\cite{Schau1987ITASSP,chan1994simple,huang2001real,li2004least,Stoica2006,
cheung2006constrained,Beck2008}.  The problem is nonconvex as the Hessian matrix of a quadratic term of
one of the constraints is not positive semi-definite. To the best of our knowledge, most of the existing methods solve an                                           approximate of~\eqref{eqn:opt-problem-LS-trans} with some information loss, e.g.,
one classical way
 is to
discard  the two quadratic constraints~\eqref{eqn:opt-problem-LS-trans-constr1}
and~\eqref{eqn:opt-problem-LS-trans-constr2}, and investigate the resulting ULS problem~\cite{li2004least,Stoica2006,Gillette08TSPL,Smith1987IJOE,smith1987closed}.
In~\cite{Beck2008}, one sufficient condition and one necessary condition are provided respectively for the CLS solution, but there is no explicit and complete characterization for global CLS solutions.

In this paper, we are concerned about solvability of the CLS problem,  the solution characterization and CLS solution uniqueness. We are also interested in exploring the solution structures for some special cases. Answers to these questions  will constitute our understanding of CLS localizability and will further facilitate the development of numerical algorithms for the exact CLS solution.

\section{Characterization of CLS Localizability}\label{section:characterization}

In this part, we present conditions on the localizability of range-difference based measurements using CLS methods.

\subsection{CLS Solution Existence}
{
First we discuss the existence and boundedness of CLS solutions. In reality we hope that
the solution set is bounded so that when some solving algorithm is implemented it would not output arbitrarily large solution. As will be shown in the following,
the boundedness of CLS solutions has connections with the Jacobian
of the TDoA measurement model~\eqref{eqn:di-squared}.

{
\begin{lemma}\label{lemma:solution-existence}
The CLS localization problem has a solution. Given the  location set 
$\{a_i:i=0,\ldots, m\}$, the CLS solution set is bounded almost surely for any set of measurements
$\{d_i: i=1,\ldots,m\}$.
\end{lemma}
}
The proof is reported in Appendix~A. To ensure a bounded CLS solution set, we introduce the following assumption.
\begin{assumption}\label{lemma:sufficient-global-minimizer}
The value of $J(x)x$ is nonzero at any nonzero $x\in \mathbb R^n$
\end{assumption}
}
In the following, we  present  a necessary and sufficient condition for a global
minimizer of the CLS problem.  The condition is developed mainly based on the method of Lagrange multipliers. The proof is reported in Appendix~B.
\begin{theorem}\label{thm:iff-condition}
Let Assumption~\ref{lemma:sufficient-global-minimizer} hold.
A vector $y\in \mathbb R^{n+1}$ is a global optimal solution of the CLS
Problem
 if and only if it falls into one of the following two cases:
 \begin{enumerate}
\item [$(i)$.] The vector $y$ satisfies $g(y) =0$ with $[y]_1>0$ and there exists a $\lambda \in \mathbb R$ such that
\begin{subequations}\label{eqn:iff-condition}
\begin{align}
(A^\top A+\lambda D)y&=A^\top b,\label{eqn:condition-1}\\
v^\top (A^\top A+\lambda D) v&\geq 0, \hbox{~for~}v\hbox{~satisfies~} v^\top Dv\leq 0 \label{eqn:indefinite-condition}
\end{align}
\end{subequations}
\item [$(ii)$.] It holds that $y=0$ and there does not exist $y'\in\mathbb{R}^{n+1}$ and $\lambda\in\mathbb{R}$ such that $y'\neq 0$, $[y']_1>0$, $g(y') =0$ and~\eqref{eqn:iff-condition} hold.
\end{enumerate}
\end{theorem}

We revisit the ULS problem, an approximate of the CLS problem, and note that the rank of $A^\top A$ dictates the number of solutions: when
$A^\top A$ is nonsingular there is a unique solution
of the ULS approximation. However, the situation is not the same for the CLS localization itself.
The nonsingularity of $A^\top A$ guarantees existence of a global solution of the CLS localization, but not necessarily uniqueness. Our further result uncovers that
uniqueness of CLS solution follows from a stricter condition, see
Theorem~\ref{thm2}. We use the following example to illustrate this fact.
 \begin{example}\label{example:1}
Consider a line-shaped (see Figure~\ref{fig:figure_TDoA}) sensor array consisting of four sensors (i.e., $m=3$) monitoring a radiating source in a 2-dimension space, i.e.,
$n=2$. The coordinates of the sensors are $a_0=0$, $a_1=[\frac{1}{\sqrt{2}},~\frac{1}{\sqrt{6}}]^\top$,
$a_2=[-\frac{1}{\sqrt{2}},~\frac{1}{\sqrt{6}}]^\top$, $a_3=[0,~-\frac{2}{\sqrt{6}}]^\top$.
We let the range-differences measurements be as follows:
$$
d_1=\frac{1}{\sqrt{3}},~~d_2=\frac{1}{\sqrt{3}},~~d_3=\frac{1}{\sqrt{3}}.
$$
Then we compute
$$
A=\left[\begin{array}{ccc}
\frac{1}{\sqrt{3}} & \frac{1}{\sqrt{2}}  & \frac{1}{\sqrt{6}}\\
\frac{1}{\sqrt{3}} & -\frac{1}{\sqrt{2}}& \frac{1}{\sqrt{6}}\\
\frac{1}{\sqrt{3}} & 0 & -\frac{2}{\sqrt{6}}
\end{array}
\right],~~~
b=\left[\begin{array}{cc}
\frac{1}{6}\\
\frac{1}{6}\\
\frac{1}{6}
\end{array}
\right].
$$
It can be verified that in this example Assumption~\ref{lemma:sufficient-global-minimizer}is satisfied
with $A^\top A=I$.
Figure~\ref{fig:graph-1}
specifies the contour lines to display
the value of $f(y)$ under the
constraints~\eqref{eqn:opt-problem-LS-trans-constr1} and
\eqref{eqn:opt-problem-LS-trans-constr2} on a plane spanned by
$[y]_2$ and $[y]_3$.  As we see in the figure,
 all points on a ring (the ring is marked with a red arrow) achieve the minimum of $f$.
This indicates that solutions of the CLS problem are not unique.
\begin{figure}[h!]
\begin{center}
\includegraphics[width=3in, height=2.7in]{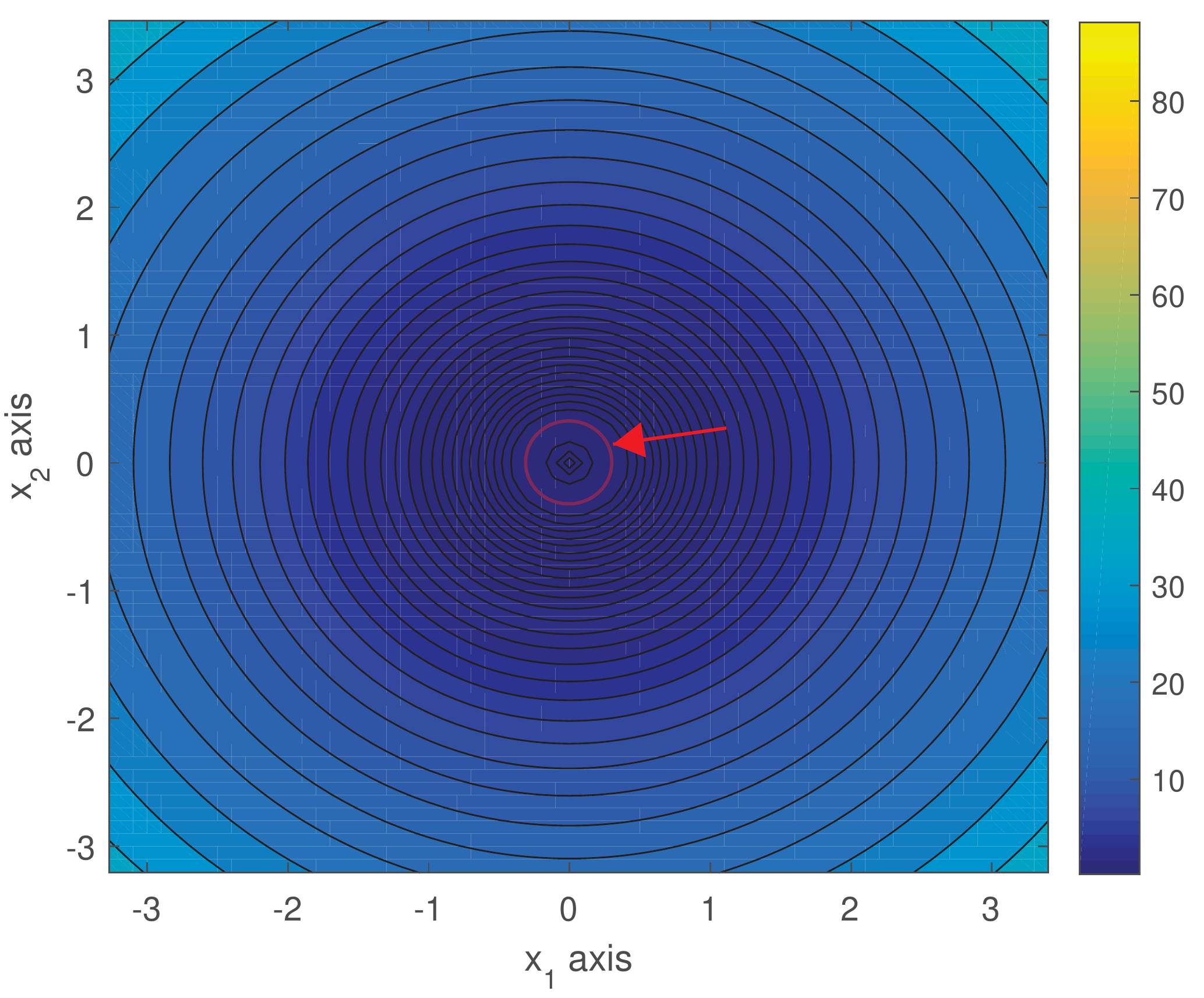}
\caption{ The contour plot of $f(y)$ on a plane spanned by
$[y]_2$ (i.e., $x_1$ axis) and $[y]_3$ (i.e., $x_2$ axis). The colormap is used to displayed different levels of $f(y)$ in the plot. All $x$'s on a ring, which is marked with a red arrow, achieve the minimum of $f$.}\label{fig:graph-1}
\end{center}
\end{figure}
\end{example}

In Theorem~\ref{thm:iff-condition}, the multiplier $\lambda$ in~\eqref{eqn:iff-condition}
 plays a critical role in affecting solutions for CLS problem.
Define the set of $\lambda$'s
satisfying~\eqref{eqn:indefinite-condition} as:
\begin{equation}\label{eqn:set_I}
\mathcal I=\{\lambda\in\mathbb R: v^\top (A^\top A+\lambda D)v\geq 0,
\forall v^\top D v \leq 0\}.
\end{equation}
Since $A^\top A\geq 0$, the inequality
$v^\top(A^\top A+\lambda D)v\geq 0$ holds for all $\lambda\leq 0$ satisfying $v^\top D v \leq0$. Thus, the set $\mathcal I$ is not empty.
The next result gives more refined descriptions of the set $\mathcal I$.
\begin{lemma}\label{lemma:interval-description}
Let Assumption~\ref{lemma:sufficient-global-minimizer} hold.
The set $\mathcal I$ can be written in the form of
$\mathcal I :=(-\infty,\lambda_u]$ with a constant $\lambda_{u}\geq 0$. Moreover,
$\lambda_u$ can be computed by solving a linear matrix inequality (LMI) problem as follows:
\begin{subequations}\label{eqn:LMI-problem-2}
\begin{align}
\mathop{\rm maximize}_{\lambda\in\mathbb R}~~~ & \lambda \\
{\rm subject~to~~~} & A^\top A+\lambda D\geq 0.
\end{align}
\end{subequations}
\end{lemma}
\noindent \textbf{Proof.~}We shall show that $\mathcal I$ is convex and therefore it is
an interval. Since $\mathcal I$ is not empty, let $\lambda_1\in \mathcal I$. For any $\lambda\in(-\infty,\lambda_1) $, we obtain
$$
v^\top (A^\top A+\lambda D)v> v^\top (A^\top A+\lambda_1 D)v\geq 0,~~\forall v^\top Dv<0.
$$
Hence, $\lambda\in \mathcal I$.
Finally, we conclude $\lambda_u\geq 0$ due to $0\in\mathcal I$.
\hfill$\blacksquare$

The above lemma, as a preliminary result of
the value range of the multipliers $\lambda$'s that satisfy~\eqref{eqn:indefinite-condition},
 depicts an interval it may locate in.
It is succeeded by a concise edition of
the CLS solution characterization as follows:
\begin{corollary}\label{corollary1}
Let Assumption~\ref{lemma:sufficient-global-minimizer} hold.
A vector $y\in \mathbb R^{n+1}$ is a global optimal solution of the CLS
Problem if and only if it falls into one of the following two cases:
 \begin{enumerate}
\item [$(i)$.] The vector $y$ satisfies $g(y) =0$ with $[y]_1>0$ and there exists a $\lambda \in \mathcal I$ such that $(A^\top A+\lambda D)y=A^\top b$.
\item [$(ii)$.] It holds that $y=0$ and there does not exist $y'\in\mathbb{R}^{n+1}$ and $\lambda\in\mathbb{R}$ such that $y'\neq0$, $[y']_1>0$, $g(y') =0$ and~\eqref{eqn:condition-1} hold.
\end{enumerate}
\end{corollary}

\noindent \textbf{Proof.~}The proof follows from
Theorem~\ref{thm:iff-condition} and the definition of the set
$\mathcal I$.

\subsection{CLS Localization Uniqueness}

In this part, on top of the existence of the CLS solutions,
we will present a sufficient and necessary condition for the solutions' uniqueness. Then we further present a finding that relates the Lagrange
multiplier to the multiplicity of the CLS solutions.

The solution characterization in Theorem~\ref{thm:iff-condition}
involves a KKT condition and a second-order condition on the Hessian matrix $A^\top A+\lambda^* D$ of a Lagrangian function.
The next result is on uniqueness of the global optimal solutions of
problem~\eqref{eqn:opt-problem-LS-trans}. We will find that
uniqueness follows
a stricter second-order condition. The proof is reported in Appendix~B.
\begin{theorem}\label{thm2}
Let Assumption~\ref{lemma:sufficient-global-minimizer} hold.
A vector $y\in \mathbb R^{n+1}$ is the unique global optimal solution of the
CLS Problem
 if and only if it falls into one of the following two cases:
 \begin{enumerate}
\item [$(i)$.]  The vector $y$ satisfies $g(y) =0$ with $[y]_1>0$ and there exists a
     $\lambda \in \mathbb R$ such that
\begin{subequations}\label{eqn:if-unique-condition}
\begin{align}
(A^\top A+\lambda D)y&=A^\top b,\label{eqn:condition-unique-1}\\
v^\top (A^\top A+\lambda D) v&> 0, \hbox{~for~}v\hbox{~satisfies~} \nonumber \\
& v^\top Dv <  0 \hbox{~and~} v^\top D y\not=0.\label{eqn:indefinite-condition-unique}
\end{align}
\end{subequations}
\item [$(ii)$.] $y=0$ and there exists no $y'\not =0$, where $[y']_1>0$, and
$\lambda\in\mathbb R$ such that
    $g(y') =0$ and~\eqref{eqn:iff-condition} hold.
\end{enumerate}
\end{theorem}
The conditions in Lemma~\ref{lemma:solution-existence} are sufficient for the existence of
a bounded global solution of the CLS problem, but not necessary.
The conditions in Theorems~\ref{thm:iff-condition} and~\ref{thm2} are both exact characterizations of some certain properties of the global solutions.

The following example demonstrates the gap between Lemma~\ref{lemma:solution-existence} and Theorems~\ref{thm:iff-condition} and~\ref{thm2}.

\begin{example}\label{example:1-contd}
We continue to consider Example~\ref{example:1}. In the example,
we have computed that $A^\top A=I$ and illustrated that the CLS
problem has bounded global solutions. Observe that
$A^\top A+D=\diag(2,0,0)$ and $A^\top b=[\frac{\sqrt{3}}{6},0,0]^\top$.
We then find that $\lambda^*=1$ and
any $y^*$, which satisfies $[y^*]_1=\frac{\sqrt{3}}{12}$ and
$\|y^*\|=\frac{\sqrt{6}}{12}$, meet the conditions~\eqref{eqn:iff-condition}.
By Theorem~\ref{thm:iff-condition}, the problem has global solutions, illustrated by
Figure~\ref{fig:graph-1}. On the other hand, for any nonzero vector $v\in\mathrm{span}
\left(\left[\begin{array}{cc}
0\\
1\\
0
\end{array}\right],
\left[\begin{array}{cc}
0\\
0\\
1
\end{array}\right]\right)$, we have $v^\top (A^\top A+\lambda^* D)v=0$.  The condition~\eqref{eqn:indefinite-condition-unique} is not met, therefore, by Theorem~\ref{thm2}, the problem's solution is not unique. Figure~\ref{fig:graph-1}
is evidence of this claim.

We next consider another example of an array consisting of four sensors. Assume that the coordinates of them are
$$a_0=a_1=0, ~a_2=[4, ~0]^\top, ~ a_3=[0,~4]^\top,$$ and that the range-difference
measurements are
$$d_1=4,~ d_2=d_3=0.$$
Then we obtain that
$$A=4I,~b=[-8, ~8, ~8]^\top.$$
\begin{figure}[h!]
\begin{center}
\includegraphics[width=3in, height=2.6in]{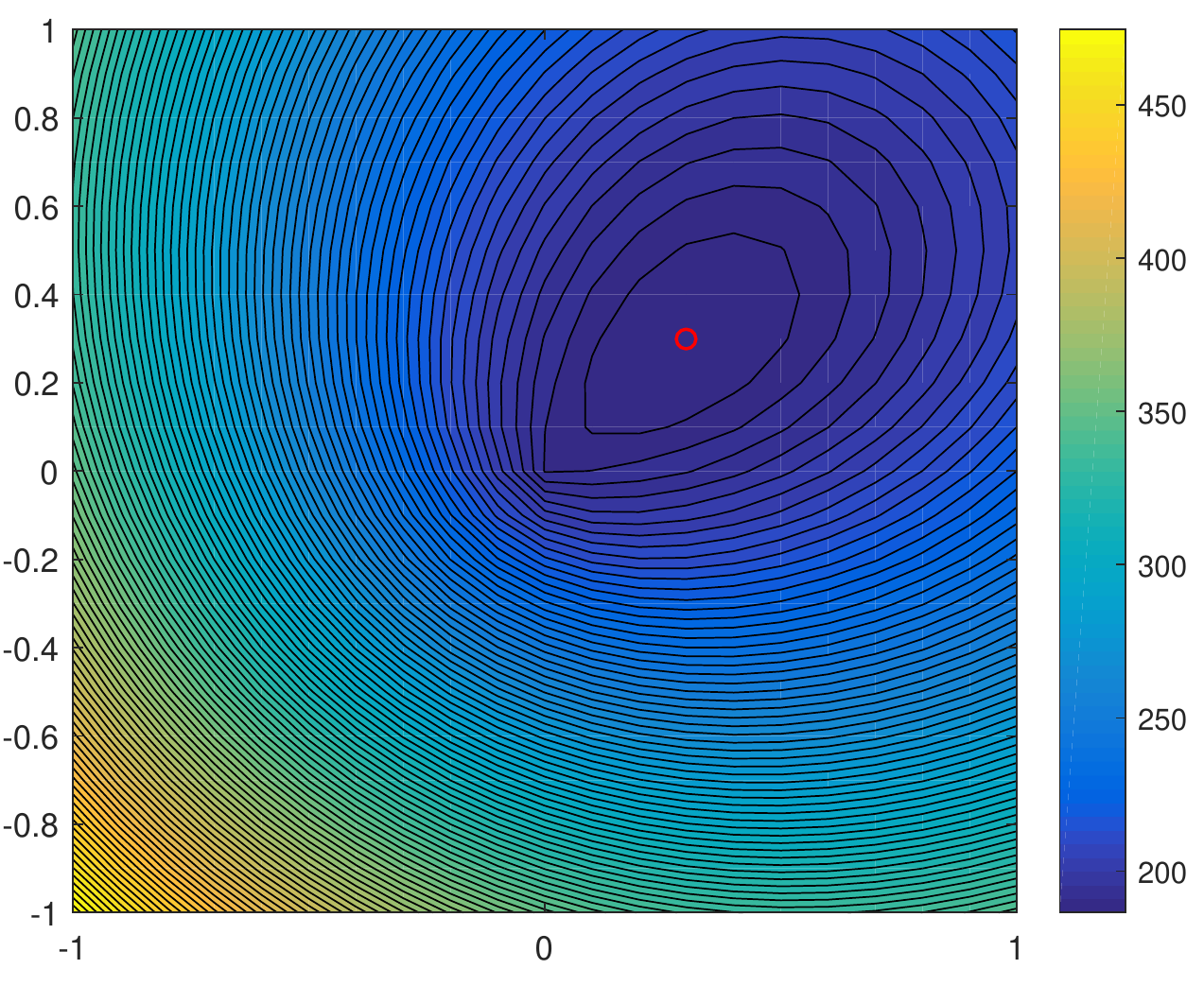}
\caption{ The contour plot of $f(y)$ on the $2$-dimensional plane. The colormap is used to displayed different levels of $f(y)$ in the plot. The point $[\frac{2-\sqrt{2}}{2},~\frac{2-\sqrt{2}}{2}]$, marked by a red circle, is
 the unique minimizer of the CLS problem in Example~\ref{example:1-contd}.}\label{fig:graph-2}
\end{center}
\end{figure}
By Theorem~\ref{thm:iff-condition}, we have $y^*=\frac{2-\sqrt{2}}{2}[\sqrt{2},~1,~1]^\top$ and $\lambda^*=-16(3+2\sqrt{2})$ satisfy~\eqref{eqn:condition-1} and~\eqref{eqn:indefinite-condition}, and therefore $\frac{2-\sqrt{2}}{2}[\sqrt{2},~1,~1]^\top$ is a
minimizer of the CLS problem. Numerical search suggests that
$\frac{2-\sqrt{2}}{2}[\sqrt{2},~1,~1]^\top$ is the unique
minimizer, see Figure~\ref{fig:graph-2}. Since $\lambda^*=-16(3+2\sqrt{2})<0$,
for any $v^\top D v<0$, we have $v^\top (A^\top A+\lambda^* D)v>0$, which theoretically supports our numerical result.
\end{example}

The following corollary is on the relations between the multiplier $\lambda$ in~\eqref{eqn:iff-condition} or~\eqref{eqn:if-unique-condition}
and the multiplicity of the CLS solutions. It unveils that
 whether a CLS problem has a single unique solution is
dictated by the position of $\lambda$ in $\mathcal I$. It directly follows from Corollary~\ref{corollary1}, Theorem~\ref{thm2} and Lemma~\ref{lemma:interval-description}.

\begin{corollary}
Let Assumption~\ref{lemma:sufficient-global-minimizer} hold.
The CLS problem has multiple global solutions
only if there exists a vector $y\in\mathbb R^{n+1}$ such that $g(y) =0$ with $[y]_1>0$
and $(A^\top A+\lambda_u D)y=A^\top b$. On the other hand,
a vector $y\in \mathbb R^{n+1}$ is a single unique global solution
if it falls into one of the following two cases:
 \begin{enumerate}
\item [$(i)$.] The vector $y$ satisfies $g(y) =0$ with $[y]_1>0$ and there exists
a $\lambda<\lambda_u$ such that $(A^\top A+\lambda D)y=A^\top b$.
\item [$(ii)$.] $y=0$ and there exists no $y'\not =0$, where $[y']_1>0$, and
$\lambda'\in\mathcal I$ such that
    $(A^\top A+\lambda' D)y=A^\top b$.
\end{enumerate}
\end{corollary}

\begin{remark}
{ The reference~\cite{More93OMS}
gives characterization of
the global minimizer of the generalized trust region problem and its uniqueness
in terms of the Lagrange multiplier.
In our paper, the existence of global solutions
to the CLS localization and solution uniqueness is derived partly based on~\cite{More93OMS}. On one hand,
the problems respectively investigated in the two papers have similar forms. The additional positivity constraint~\eqref{eqn:opt-problem-LS-trans-constr2} require
 more elaborate analysis to derive conditions for CLS solution and its uniqueness.
On the other hand, the conditions in our paper are less strict
 than those of
~\cite{More93OMS} for the generalized trust region problem  since~adding \eqref{eqn:opt-problem-LS-trans-constr2} into the constraints defines a narrowed feasible set.  }
\end{remark}

\subsection{Literature Comparison and Discussions}
In this part, we present comparison between our aforementioned results and
the related ones in literature.

First note that a characterization of the global solutions to the CLS problem
was presented in~\cite{Beck2008}, giving a sufficient condition and a necessary condition respectively.
There is a high degree of similarity between the CLS problem
and generalized trust region problems (GTRS)~\cite{More93OMS},
a class of problems minimizing a quadratic function subject to a quadratic
equality constraint that has been reviewed by us in above. A characterization of the global solution was obtained
in~\cite{More93OMS}. In the following remarks, we will compare the
results of this paper with those of~\cite{Beck2008} and~\cite{More93OMS}, respectively.

In~\cite{Beck2008}, a sufficient condition, i.e.,~\cite[Lemma~3.1]{Beck2008}, and a necessary
condition, i.e.,~\cite[Theorem~3.1]{Beck2008}, for solutions of the CLS problem were
presented. The gaps in between them two are discussed as follows:
\begin{inparaenum}[(i).]
\item  the sufficient condition requires $A^\top A+\lambda D$ to have  no negative eigenvalues while the necessary one is relaxed,
    allowing
$A^\top A+\lambda D$ to have at most one negative eigenvalue;
\item  the possibility of $y_{\rm CLS}^*=0$ is excluded from the sufficient condition while not excluded in the necessary one.
\end{inparaenum}
These gaps are not
trivial in the sense that there may exist cases where a global minimizer $y^*_{\rm CLS}$ fails to satisfy the sufficient condition
as well as cases where the necessary condition does not lead to
a solution.
In Example~\ref{example:3-example1-contd}, we demonstrate two cases
where the gaps exist.
\begin{example}[Example~\ref{example:1-contd} Cont'd]\label{example:3-example1-contd}
In this example, we continue to study the cases in Example~\ref{example:1-contd}.
First, we consider the same scenarios to the first case of  Example~\ref{example:1-contd}. We have shown that $\lambda^*=1$ and
any $y^*$ that satisfies $[y^*]_1=\frac{\sqrt{3}}{12}$ and
$\|y^*\|=\frac{\sqrt{6}}{12}$ meet the conditions~\eqref{eqn:iff-condition}.
Then, we have $A^\top A+\lambda^* D=\diag(2,0,0) $. The positive semidefiniteness of $A^\top A+\lambda^* D$ indicates the conservativeness of the necessary condition of~\cite[Lemma~3.1]{Beck2008}.

We next consider the second scenario of Example~\ref{example:1-contd}.
It has been computed in Example~\ref{example:1-contd}
 that $y=\frac{2-\sqrt{2}}{2}[\sqrt{2},~1,~1]^\top$
  is an optimal solution in this case(see Figure~\ref{fig:graph-2}) and
$\lambda=-16(3+2\sqrt{2})$ satisfies~\eqref{eqn:condition-1} and~\eqref{eqn:indefinite-condition}. However, simple
computation reveals that, having a negative eigenvalue, $A^\top A+\lambda D>0$
 does not hold.
It suggests
conservativeness of the sufficient condition of~\cite[Lemma~3.1]{Beck2008}.
\end{example}

The CLS problem has a close relation with the GTRS problem as they have a similar form. The form of the GTRS problem is as follows~\cite{More93OMS}:
\begin{subequations}\label{eqn:opt-problem-GTRS}
\begin{align}
\hbox{\textbf{(GTRS)}:}~~~~~~
\mathop{\rm minimize~}\limits_{y\in\mathbb R^{n+1}} & f(y)\label{eqn:opt-problem-GTRS-obj}\\
\mathop{\rm subject~to~} & g(y)=0.\label{eqn:opt-problem-GTRS-constr1}
\end{align}
\end{subequations}
Compared to the CLS problem, the GTRS
does not contain the constraint $[y]_1\geq 0$.
By~\cite[Theorems 3.2]{More93OMS}, a
 vector $y^*_{\rm GTRS}\in \mathbb R^{n+1}$ is a global minimizer
 if and only if $g(y^*_{\rm GTRS}) =0$ and there exists a $\lambda^*_{\rm GTRS} \in \mathbb R$ such that~\eqref{eqn:condition-1}
 and  $A^\top A+\lambda^*_{\rm GTRS} D\geq 0$ hold. Moreover by~\cite[Theorems 3.3]{More93OMS},
 in case of $A^\top A+\lambda^*_{\rm GTRS} D>0$, $y^*_{\rm GTRS}$ is the unique solution.

A comparative study to the characterization of solutions to
the GTRS and the CLS problems reveals some close observations.
First, Theorems~\ref{thm:iff-condition} and~\ref{thm2} of our paper and
Theorems 3.2 and 4.1 of~\cite{More93OMS} shared similar KKT conditions
 because
the CLS and the GTRS problems have similar forms and it
is allowed to have
analytic
characterization of the solutions using the same Lagrangian function.
On the other hand, these problems have different second-order conditions
because
for the CLS problem
the constraint~\eqref{eqn:opt-problem-LS-trans-constr2}
gets rid of the lower napple\footnote{A double cone is a quadratic surface.
Each single cone placing apex to apex is called a napple.}
of the double cone $\{y:y^\top D y=0\}$  and
this allows negative definiteness of
the Hessian matrix of~\eqref{eqn:Lagrangian-function} in some subspace.

\section{Structured CLS Localizations}\label{section:solution_algorithm}

{
The characterization of CLS global minimizers established in
Theorem~\ref{thm:iff-condition}
paves the way for us to locate the minimizers by resorting to
seeking for feasible $\lambda$'s.
A viable way for doing so is to verify the
conditions $(i)$--$(ii)$ of Theorem~\ref{thm:iff-condition} in order:
first search multipliers $\lambda$'s
satisfying~\eqref{eqn:indefinite-condition}; substituting each
$\lambda$ into~\eqref{eqn:condition-1}, then
solve a solution of~\eqref{eqn:condition-1} and check the feasibility
of~\eqref{eqn:opt-problem-LS-trans-constr1} and~\eqref{eqn:opt-problem-LS-trans-constr2}.
Besides, we can do better in some special cases, where a global minimizer of the CLS problem can be computed easily. This is done on the basis of
the theory developed in the previous section.
In what follows, we will proceed to present how it is realized.

To guarantee that the CLS problem exists solutions, we assume that \emph{Assumption~\ref{lemma:sufficient-global-minimizer} holds in the rest of
 this section}, and  will elaborate how $\lambda_{\rm CLS}^*$ is positioned
in different situations.

\subsection{Technical Preparations}
Define the following set
\begin{equation}\label{eqn:set_I2}
\mathcal I_{1}=\{\lambda\in\mathbb R: A^\top A+\lambda D> 0\}.
\end{equation}
Obviously $\mathcal I_1\subset \mathcal I$ and
due to the indefiniteness of $D$ as well as Assumption~\ref{lemma:sufficient-global-minimizer} the set $\mathcal I_1$ is a non-empty bounded interval . This interval can be formally represented by  $(\lambda_l,\lambda_u)$, where $\lambda_l\in\mathbb R$ and
is the solution of the following LMI problem:
\begin{align*}
\mathop{\rm minimize}_{\lambda\in\mathbb R}~~~ & \lambda \\
{\rm subject~to~~~} & A^\top A+\lambda D\geq 0.
\end{align*}
By~\cite[Theorem~5.3]{More93OMS}, the closure of $\mathcal I_1$, denoted by
$\overline{\mathcal I_1}:=[\lambda_l,\lambda_u]$, has the following property:
$$\overline{\mathcal I_1}=\{\lambda\in\mathbb R:A^\top A+\lambda D\geq 0\},$$
and $A^\top A+\lambda D\geq 0$ is singular for $\lambda=\lambda_l \hbox{~and~}\lambda_u$.
In the presence of the specific form of $D$ given in~\eqref{eqn:def-D},
we further reach claims that $\lambda_l<0$ and that, as an eigenvalue of $A^\top A+\lambda_l D$, zero has a  geometric multiplicity $M_0=1$. This claim can be proved in the following way. If $M_0$ were greater than $1$, we consider two distinct
subspaces $\mathcal U$ and $\mathcal V$, defined as
$\mathcal U:=\{u\in\mathbb R^{n+1}: (A^\top A+\lambda_l D)u=0\}$
and $\mathcal V:=\{v\in\mathbb R^{n+1}: (D+I)v=0\}$.
By calculation, the dimensions of $\mathcal U$ and $\mathcal V$ satisfy
${\rm dim}(\mathcal U)=M_0$ and
${\rm dim}(\mathcal V)=n$.
By the Grassmann's formula~\cite[2.18~Corollary]{Giaquinta07Mathematicalanalysis},
\begin{align}\label{eqn:dim_eqn}
{\rm dim}(\mathcal{U})+{\rm dim}(\mathcal{V})={\rm dim}(\mathcal{U}\cup \mathcal{V})+{\rm dim}(\mathcal{U}\cap\mathcal{V}).
\end{align}
Since ${\rm dim}(\mathcal U\cup \mathcal V)\leq n+1$, we have that
${\rm dim}(\mathcal U\cap \mathcal V)\geq M_0-1\geq 1$. However,
for any $v\in\mathcal V\setminus\{0\}$, $v^\top (A^\top A+\lambda_l D)v=
v^\top A^\top A v-\lambda_l \|v\|^2>0$. Hence $v\not\in\mathcal U$ and
$\mathcal U\cap \mathcal V=\{0\}$, which reaches a contradiction.

Letting $\mathcal I_2:=(-\infty,\lambda_l)$, we have
that $\mathcal I=\overline{\mathcal I_1}\cup \mathcal I_2$ and that, for any $\lambda\in\mathcal I_2$, the matrix $A^\top A+\lambda D$ is nonsingular, or to be more specific, it has exactly one negative eigenvalue and $n$ positive eigenvalues.
We call \textit{$\mathcal I_1$ the positive-definite interval (PDI)} for an optimal
Lagrangian multiplier and call \textit{$\mathcal I_2$ the indefinite interval (IDI)}, and call $\lambda_l$ and
$\lambda_u$ \textit{the left singular point (LSP) and right singular point (RSP)}, respectively.
We use
$y(\lambda)$ to denote a solution to~\eqref{eqn:condition-1},
where the argument $\lambda$ in $y(\lambda)$ reminds that it is a solution
under a given $\lambda\in\mathcal \overline{\mathcal I}_1\cup \mathcal I_2$,
and denote
\begin{equation}\label{eqn:func_g}
h(\lambda):=g(y(\lambda)).
\end{equation}
We will show that the function $h(\lambda)$ exhibits some nice properties and develop analysis
in some special cases on the ground of them. In what follows, we will present
these cases one by one. Let us first recall notions that
have been defined and will be used frequently
in the sequel.

\noindent{\bf Notations.} The functions $f$, $g$ and $h$ are defined in~\eqref{eqn:opt-problem-LS-trans} and~\eqref{eqn:func_g}, respectively. The parameter $b$ is defined in~\eqref{eqn:di-squared}. The sets $\mathcal I$
and $\mathcal I_1$ are defined in~\eqref{eqn:set_I} and~\eqref{eqn:set_I2}.
We have knowledge of $\mathcal I_1$ that it is an open and bounded interval.
The set $\overline{\mathcal I_1}$ is the closure of $\mathcal I_1$, and
$\lambda_l$ and $\lambda_u$ are the left and right end points of $\overline{\mathcal I_1}$, respectively.
The set $\mathcal I_2$ is $\mathcal I_2=\mathcal I\setminus \overline{\mathcal I_1}$.
The variables $y^*_{\rm GTRS}\in \mathbb R^{n+1}$ and $\lambda^*_{\rm GTRS} \in \mathbb R$ relate to the GTRS problem~\eqref{eqn:opt-problem-GTRS}.
They  satisfy
$g(y^*_{\rm GTRS}) =0$ and ~\eqref{eqn:condition-1} and  $A^\top A+\lambda^*_{\rm GTRS} D\geq 0$---in this situation $y^*_{\rm GTRS}$ is a global minimizer
of the GTRS problem. Similarly $y^*_{\rm CLS}\in \mathbb R^{n+1}$ and $\lambda^*_{\rm CLS} \in \mathbb R$ are two variables that meet the conditions in Theorem~\ref{thm:iff-condition}.

\begin{table}[!htb]
	\centering
	\caption{{Key Notations in Sections IV.B-IV.D.}} \label{tabel3}
		\resizebox{\columnwidth}{!}{%
	\begin{tabular}{||c |c ||}
		\hline
		Notation & Description \\
		\hline
$\mathcal I$  & interval of the Lagrangian multiplier defined in~\eqref{eqn:set_I}\\
        \hline
$\mathcal I_1$ & positive-definite interval (PDI) defined in~\eqref{eqn:set_I2}\\
        \hline
$\overline {I_1}$ & closure of $\mathcal I_1$ \\
        \hline
$\mathcal I_2$ & indefinite interval (IDI), $\mathcal I_2=\mathcal I\setminus \overline{\mathcal I_1}$\\
        \hline
$\lambda_l$ ($\lambda_u$) & left singular point (LSP)(right singular point (RSP)) of $\overline {I_1}$\\
        \hline
$y^*_{\rm GTRS}$ &  global minimizer of the GTRS problem~\eqref{eqn:opt-problem-GTRS}\\
        \hline
$y^*_{\rm CLS}$, $\lambda^*_{\rm CLS}$ & variables that satisfy the conditions in Theorem~\ref{thm:iff-condition}\\
		\hline
	\end{tabular}}
\end{table}

 \subsection{The Case of Collinear Source Position ($b=0$)}
We consider the case of $b=0$. This case is special in the sense that it allows us to exactly obtain the solution immediately.
When $b=0$, one can, in a straightforward manner, derive the following result (The proof is skipped because of its simplicity.):
\begin{lemma}\label{lemma:h-monotonicity}
If $b=0$, we have that $h(\lambda)=0$ for $\lambda\in \mathcal I_1\cup \mathcal I_2$.
\end{lemma}
It can be readily validated that $y=0$ and any $\lambda \in\mathcal I_1\cup \mathcal I_2$ satisfies condition $(i)$ of Theorem~\ref{thm2}. Hence, in this case  we conclude that
the origin is the unique solution to the CLS problem.
Other facts
that we would like to point out are that the uniqueness of $\lambda_{\rm CLS}^*$ is not needed
for the uniqueness of $y_{\rm CLS}^*$ and that when $b=0$ the origin is also the unique minimizer of $f(y)$.

The position of the CLS solution for $b=0$ can also be inferred geometrically. By~\eqref{eqn:di-squared} we have $\|a_i\|=d_i$ for all sensor $i$. The geometry in Figure~\ref{fig:figure_TDoA} alludes that the best estimate of the radiating source's position is collinear with $a_0$ and any $a_i$. To make sure that Assumption~\ref{lemma:sufficient-global-minimizer} holds, $a_0$ and $a_1, \ldots, a_m$ cannot be in a line. As a consequence, the origin must be the unique best position estimate.

Next we turn to consider the case of $b\not=0$, which needs to be treated by more refined analysis.

\subsection{Searching CLS solution over PDI}

In case of $b\not=0$, $h(\lambda)$ is strictly monotonic on $\mathcal I_1$. The property was originally derived in~\cite[Theorem 5.2]{More93OMS} and is rephrased as follows:
\begin{lemma}{\cite[Theorem 5.2]{More93OMS}}\label{lemma:h-monotonicity}
Assume that $b\not =0$. The function $h(\lambda)$ is strictly decreasing on $\mathcal I_1$.
\end{lemma}

We revisit the GTRS problem~\eqref{eqn:opt-problem-GTRS}. Theorem 3.2 of~\cite{More93OMS} reads that
 a vector $y_{\rm GTRS}^*\in \mathbb R^{n+1}$ is a global minimizer of the
 GTRS problem
 if and only if $g(y_{\rm GTRS}^*) =0$ and there exists a $\lambda_{\rm GTRS}^* \in \overline{\mathcal I_1}$ such that~\eqref{eqn:condition-1}
 holds.
It can be further
concluded that, if in addition $[y_{\rm GTRS}^*]_1\geq 0$ holds, $y_{\rm GTRS}^*$  is also a global minimizer of the CLS problem.

Based on the above conclusions, we can first
search $\lambda_{\rm GTRS}^*$  over ${\mathcal I_1}$, if any, and then
compute $y_{\rm GTRS}^*$.
Lemma~\ref{lemma:h-monotonicity} reveals that
 $h(\lambda)$ is monotonically decreasing on $\mathcal I_1$.
If $A^\top A+\lambda_{\rm GTRS}^* D>0$,
a bisection algorithm can be applied for locating $\lambda_{\rm GTRS}^*$~\cite{Beck2008} by
evaluating the sign of $h(\lambda)$. Then
$y_{\rm GTRS}^*=y(\lambda_{\rm GTRS}^*)=(A^\top A+\lambda_{\rm GTRS}^* D)^{-1}A^\top b$ can be computed naturally.
By examining whether $[y_{\rm GTRS}^*]_1$ is positive or not,
one can determine whether $y_{\rm GTRS}^*$ is a global minimizer to~the CLS problem. To be precise, $y_{\rm GTRS}^*$ is a global solution to~\eqref{eqn:opt-problem-LS-trans} in the presence of $[y_{\rm GTRS}^*]_1>0$.
For the case of $A^\top A+ \lambda_{\rm GTRS}^* D$ being positive semidefinite and
singular, one needs other analysis for locating $\lambda_{\rm GTRS}^*$, which will be treated in the following subsection.

\subsection{Searching CLS solution on Singular Points}\label{sectoin:positive-semidefinite-case}

If $A^\top A+\lambda_{\rm GTRS}^* D$ is positive semidefinite and
singular, the sign of $h(\lambda)$ does not change on $\mathcal I_1$. In this situation,
by~\cite[Theorem 5.4]{More93OMS}, the following claims can be established:
\begin{enumerate}
\item[$(i)$.]
 The value of $\lambda_{\rm GTRS}^*$ is pushed towards an endpoint, either $\lambda_l$ or $\lambda_u$, of $\mathcal I_1$. More precisely, $\lambda_{\rm GTRS}^*=\lambda_u$ in case of $h(\lambda)>0$ on
$\mathcal I_1$ and $\lambda_{\rm GTRS}^*=\lambda_l$ otherwise.
\item[$(ii)$] The limit $\lim_{\lambda\to \lambda_{\rm GTRS}^*}h(\lambda)$ exists and
$\lim_{\lambda\to \lambda_{\rm GTRS}^*}y(\lambda)=y_*$ for some
$y_*\in\mathbb R^{n+1}$.
\end{enumerate}
It is pointed out by the reference~\cite{More93OMS} that $y_*$ is not necessarily
a solution to~\eqref{eqn:opt-problem-GTRS}.  In general,
it requires a subtle treatment to compute a $y_{\rm GTRS}^*$. Besides what
has been concluded in~\cite{More93OMS}, in the presence of the
specific form of the matrix $D$, see~\eqref{eqn:def-D}, we can get access to better properties related to $h(\lambda)$ and $y_{\rm GTRS}^*$, which make it possible to
determine the sign of $[y_{\rm GTRS}^*]_1$ even prior to solving the exact value.
By knowing the sign of $[y_{\rm GTRS}^*]_1$, we can determine whether $y_{\rm GTRS}^*$  is also a CLS solution.
The whole computation procedure will be detailed for the rest of this subsection.

First we focus on how the sign of $h(\lambda)$ affects that of $[y_{\rm GTRS}^*]_1$.
The analysis is divided into two parts on the ground of either $h(\lambda)>0$ or
$h(\lambda)<0$.

\subsubsection{The case $h(\lambda)<0$ for $\lambda\in \mathcal I_1$}
In this case, $\lambda_{\rm GTRS}^*=\lambda_l$. We choose $z^-$ with $\|z^-\|=1$
such that
$$(A^\top A+\lambda_l D)z^-=0.$$
Since $0$ is an eigenvalue of $A^\top A+\lambda_lD$ with geometric multiplicity $1$, $z^-$ is unique up to sign.
Then $y_{\rm GTRS}^*= y_*+\alpha z^-$ for some $\alpha\in\mathbb R$ (Recall that $\lim_{\lambda\to \lambda_{\rm GTRS}^*}y(\lambda)=y_*$.).
Here $\alpha$  is chosen as either of the roots of the following
quadratic equation (see~\cite[Section~5]{More93OMS}):
\begin{equation}\label{eqn:quadratic-eqn}
g(z^-) \alpha^2 +2y_*^\top D z^- \alpha +g(y_*)=0.
\end{equation}
The roots have the following property. The proof can be found in Appendix~C.
\begin{lemma}\label{lemma:distinct-solutions}
The quadratic equation~\eqref{eqn:quadratic-eqn} has two distinct roots.
\end{lemma}

To distinguish the two roots of~\eqref{eqn:quadratic-eqn},
 without loss of generality, we denote them by $\alpha_{1}^-$
and $\alpha_2^-$, where $\alpha_1^->\alpha_2^-$, and have the following claim.
\begin{lemma}\label{lemma:sign-of-y-lambda-l}
One of $[y_*+\alpha_1^-z^-]_1$ and $[y_*+\alpha_2^-z^-]_1$
is negative and the other is positive, i.e., $[ y_*+\alpha_1^-z^-]_1[ y_*+\alpha_2^-z^-]_1<0$.
\end{lemma}
The proof is given in Appendix C.

To summarize, when $h(\lambda)<0$ for $\lambda\in \mathcal I_1$, we have $\lambda_{\rm CLS}^*=\lambda_l$ and can solve the CLS problem by resorting to Lemma~\ref{lemma:sign-of-y-lambda-l}.
We obtain $y_*$ through $\lim_{\lambda\to \lambda_{l}}y(\lambda)=y_*$ and choose a
unit eigenvector $z^-$ of $A^\top A+\lambda_lD$ with respect to
eigenvalue $0$.
Then we compute $\alpha_1$ and $\alpha_2$ by solving~\eqref{eqn:quadratic-eqn}.
Lemma~\ref{lemma:sign-of-y-lambda-l} guarantees that
either $y_*+\alpha_1z^-$ or $y_*+\alpha_2z^-$ is a global
minimizer to~\eqref{eqn:opt-problem-LS-trans}.
\begin{example}\label{example:4}
Consider a sensor array consisting of five sensors (i.e., $m=4$) monitoring a radiating source in a 3-dimension space, i.e.,
$n=3$. The coordinates of the sensors are $a_0=0$, $a_1=[-1, 0,0]^\top$,
$a_2=[1,-1,0]^\top$, $a_3=[1,1,1]^\top$ and
$a_4=[0,1,0]^\top$.
We set the range-differences measurements to be:
$$d_1=-1,~d_2=1,~d_3=0,~d_4=-1.$$
Then we compute
$$
A=\left[\begin{array}{cccc}
-1 & -1  & 0 & 0\\
1 & 1& -1 & 0\\
0 & 1 & 1 & 1\\
-1 & 0 & 1 &0
\end{array}
\right],~~~
b=\left[\begin{array}{cc}
0\\
0.5\\
1.5\\
0
\end{array}
\right].
$$
It can be verified that $A^\top A$ is nonsingular so
the problem has solutions by Lemma~\ref{lemma:solution-existence}.
We obtain $\mathcal I_1 = (-1,0.2679)$ using ${\rm cvx}$ in Matlab.
The Matlab
simulation reads
that $h(\lambda) < 0$ and it is strictly decreasing for $\lambda\in\mathcal I_1$.
By the result in this part, we have $\lambda_{\rm GTRS}^*=\lambda_l=-1$ and
$z^-=\frac{1}{\sqrt{6}}[2,-1,1,0]^\top$.
Figure~\ref{fig:graph-example4} plots the values of $y(\lambda)$ for
$\lambda\in(-1,1]$ and reads that $y(\lambda)$ converges to $y_{*}\approx [-0.25,0.5,0,0.5]^\top$ when $\lambda$ approaches $-1$. By solving~\eqref{eqn:quadratic-eqn}, we obtain $\alpha_1^{-}=2\sqrt{21}-\sqrt{6}$
and $\alpha_2^{-}=-2\sqrt{21}-\sqrt{6}$. Then we find that
$y_*+\alpha_1^-z^-= [2\sqrt{14}-2,-\sqrt{14}+4,\sqrt{14},4]^\top $ and
$y_*+\alpha_2^-z^-= [-2\sqrt{14}-2,\sqrt{14}+4,-\sqrt{14},4]^\top $, which verifies
Lemma~\ref{lemma:sign-of-y-lambda-l}. Finally, we conclude that $y_*+\alpha_1^-z^-$ is a solution.

\begin{figure}[h!]
\begin{center}
\includegraphics[width=3in, height=2.6in]{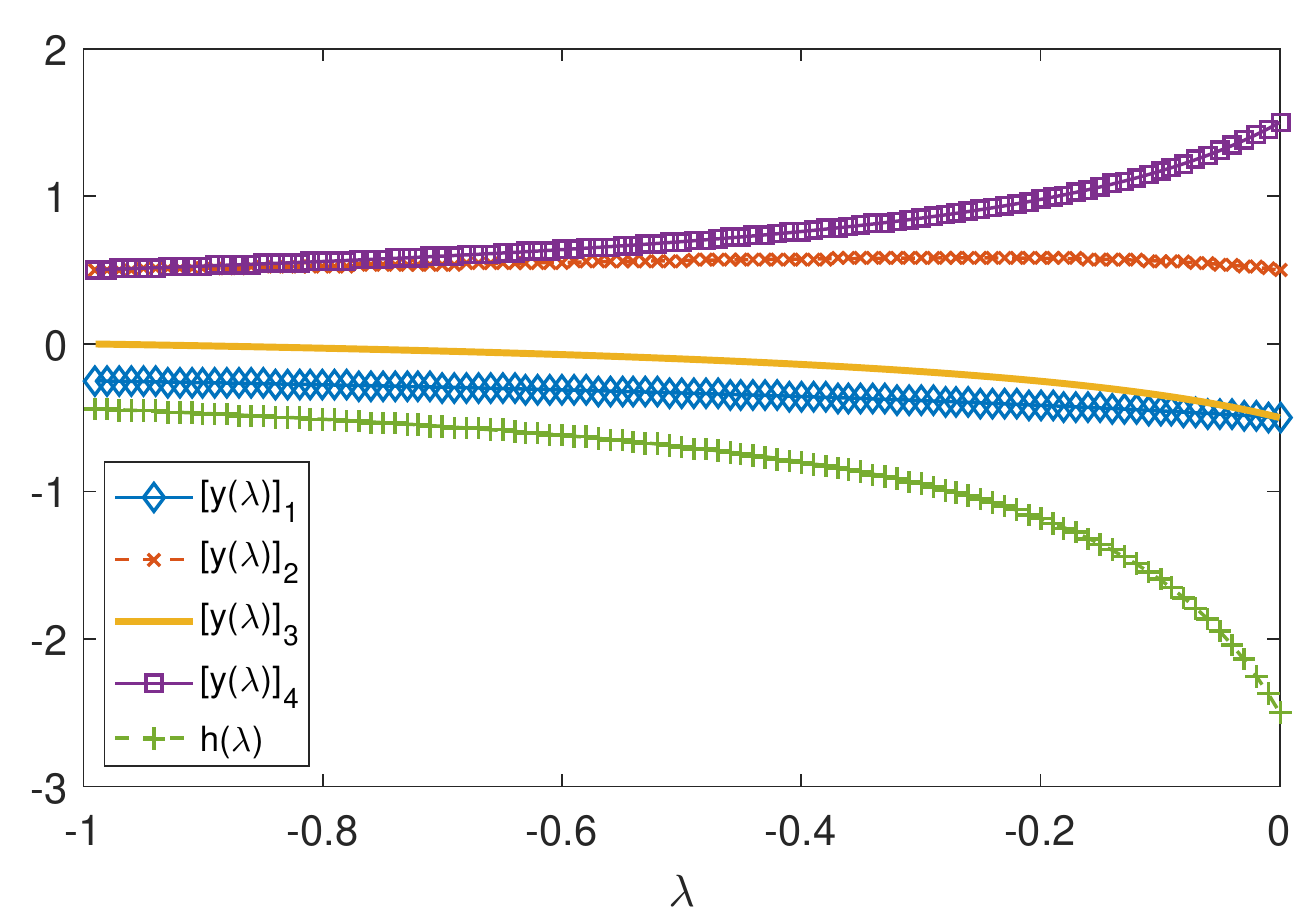}
\caption{In Example~\ref{example:4}, the Matlab simulation reads
that $h(\lambda)<0$ is strictly decreasing when $\lambda\in(-1,0]$ and  $y(\lambda)$ converges to $y_{*}\approx [-1,2,0,2]^\top$ when $\lambda$ approaches $-1$.}\label{fig:graph-example4}
\end{center}
\end{figure}
\end{example}

\subsubsection{The case of $h(\lambda)>0$ for $\lambda\in \mathcal I_1$}
In this case, $\lambda^*_{\rm GTRS}=\lambda_u$. Similarly, we choose a vector $z^+$ with $\|z^+\|=1$
such that $(A^\top A+\lambda_u D)z^+=0.$
Then $y^*_{\rm GTRS}= y_*+\alpha z^+$, where $\alpha$ is a solution of the
quadratic equation:
\begin{equation}\label{eqn:quadratic-eqn-2}
g(z^+) \alpha^2 +2y_*^\top D z^+ \alpha +g(y_*)=0.
\end{equation}
Without loss of generality, we denote the roots (which is not necessarily distinct) by $\alpha_{1}^+$
and $\alpha_2^+$, where $\alpha_1^+\geq \alpha_2^+$. The following result holds.
\begin{lemma}\label{lemma:sign-of-y-lambda-u}
The values of $[y_*+\alpha_1^+z^+]_1$ and $[y_*+\alpha_2^+z^+]_1$
are simultaneously negative or positive, i.e., $[ y_*+\alpha_1^+z^+]_1[y_*+\alpha_2^+z^+]_1>0$. Moreover, it is true that
$[y(\lambda)]_1$, $[y_*]_1$, $[y_*+\alpha_1^+z^+]_1$
and $[y_*+\alpha_2^+z^+]_1$
have the same sign for any $\lambda\in\mathcal I_1$.
\end{lemma}
The proof is given in Appendix C.

To summarize, when $h(\lambda)>0$ for $\lambda\in \mathcal I_1$,
$\lambda^*_{\rm GTRS}=\lambda_u$.
We compute $y_*$ through $\lim_{\lambda\to \lambda_{u}}y(\lambda)=y_*$ and choose a unit eigenvector $z^+$ of $A^\top A+\lambda_uD$ with respect to the
eigenvalue $0$. Then $y_{\rm GTRS}^*=y_*+\alpha_2^+z^+$ or $y_*+\alpha_2^+z^+$.
To decide whether $y_{\rm GTRS}^*$ is a global minimizer of the CLS problem, we
need to identify the sign of $[y_{\rm GTRS}^*]_1$.
Lemma~\ref{lemma:sign-of-y-lambda-u} allows us to do this by simply
examining the sign of any $[y(\lambda)]_1$ for $\lambda\in \mathcal I_1$.
When $0\in \mathcal I_1$,
a simple means is to check the sign of $[y(0)]_1=[(A^\top A)^{-1}A^\top b]_1$.
If $[(A^\top A)^{-1}A^\top b]_1>0$, then $y_{\rm GTRS}^*$ is a CLS solution;
otherwise it is not.
\begin{example}[Example~\ref{example:1} Cont'd]\label{example:5}
We continue to consider Example~\ref{example:1}. In this example, $\mathcal I_1=(-1,1)$. Our Matlab simulation reads
that $h(\lambda)>0$ is strictly decreasing for $\lambda\in\mathcal I_1$, see Figure~\ref{fig:graph-example5}.
By the result in this part, we have $\lambda_{\rm GTRS}^*=\lambda_u=1$, which is consistent with the result in Example~\ref{example:1-contd}. Moreover, Figure~\ref{fig:graph-example5} plots the values of $y(\lambda)$ for
$\lambda\in[0,1)$ and reads that $y(\lambda)$ converges to $y_{*}\approx [0.145,0,0]^\top$ when $\lambda$ approaches $1$, and that
$[y_{*}]_1$ and $[y(\lambda)]_1$'s are all positive numbers.
we choose $z^+=[0,1,0]^\top$ since $(A^\top A+\lambda_u D)z^+=0$. Solve
the equation $g(y_*+\alpha z^+)=0$, and we obtain that $\alpha\approx \pm 0.145$,
hence $y_{\rm GTRS}^*=y_*+\alpha z^+\approx [0.145, \pm 0.145, 0]^\top$. The first
elements of the two $y_{\rm GTRS}^*$'s are both positive. So far the Matlab simulation has validated Lemma~\ref{lemma:sign-of-y-lambda-u}. The optimal solution numerically simulated in Matlab rather meets the ones computed  theoretically in  virtue of Theorem~\ref{thm:iff-condition}.

\begin{figure}[h!]
\begin{center}
\includegraphics[width=3in, height=2.6in]{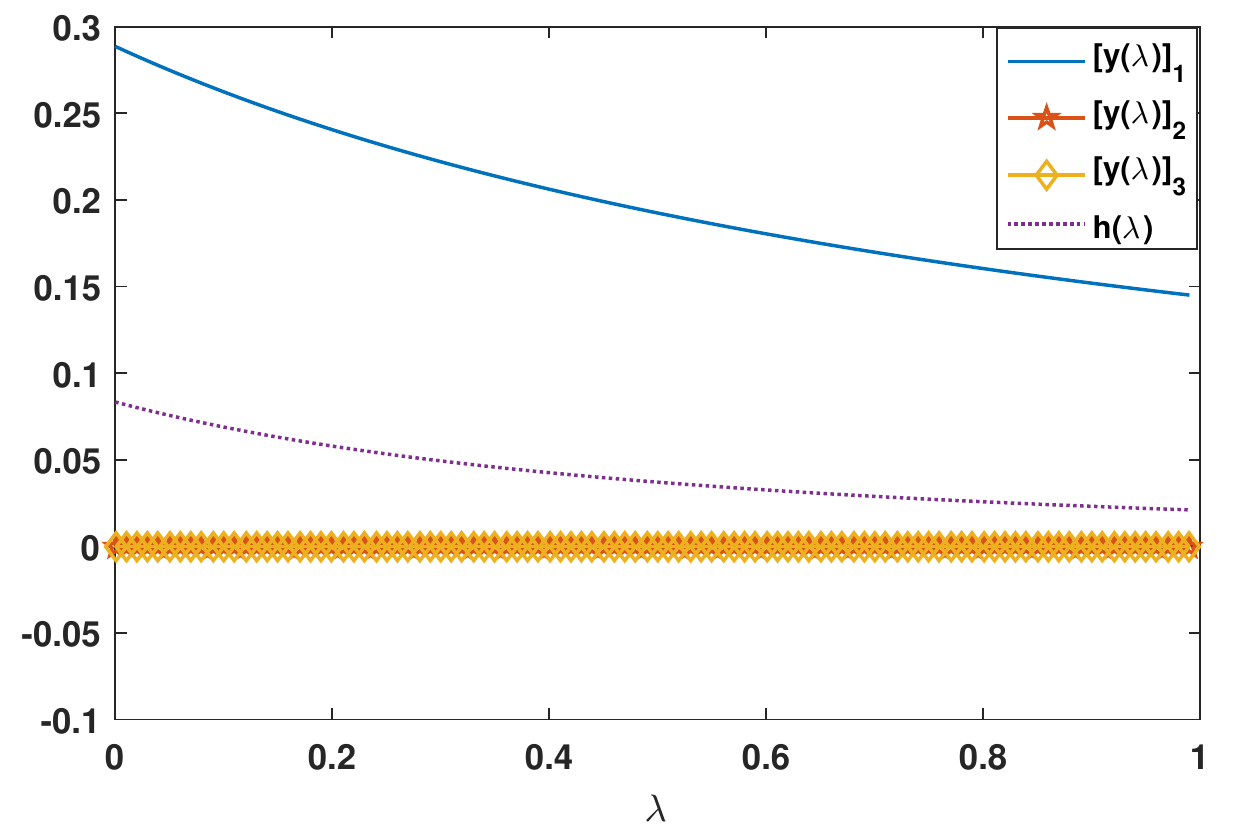}
\caption{In Example~\ref{example:5}, the Matlab simulation reads
that $h(\lambda)>0$ is strictly decreasing when $\lambda\in[0,1)$ and  $y_{*}\approx [0.145,0,0]^\top$. In addition, it can be observed that
$[y_{*}]_1$ and $[y(\lambda)]_1$'s are all positive numbers.}\label{fig:graph-example5}
\end{center}
\end{figure}
\end{example}

{
	\section{Numerical Simulation}
	We search optimal solutions of the CLS problems by utilizing the results we develop in the paper. The methods compared with ours are as follows: (a) ULS: unconstrained least squares solution~\cite{Gillette08TSPL}; (b) SDP-I: inner-product semidefinite relaxation algorithm~\cite{xu2010reduced}; (c) CFS: the classical TDOA closed-form solution by Chan and Ho~\cite{chan1994simple}; (d) SUM-MPR: modified polar representation solved by the successive unconstrained minimization approach~\cite{sun18tsp}.
	
	\emph{Example 6.} Let the reference sensor sit at the origin and the other four sensors locate in 
	\begin{align*}
	&a_1=[-1 ~ 1]^\top,~a_2=[-1 ~ 4]^\top, \\
	&a_3=[-4 ~ 6]^\top,~a_4=[-6 ~ 7]^\top. 
	\end{align*}
	The coordinates of the source is set as $[-5 ~ 2]^\top $. The range-difference measurements $d_i,i=1,\ldots,4$ are corrupted by i.i.d. Gaussian noises $r_i \sim \mathcal N(0,\sigma^2)$. For each given $\sigma$, we run $N=1000$ times and calculate the average squared error
	$\frac{1}{N}\sum_{i=1}^{N}\left\| {\tilde x^*(\omega_i) - x^*} \right\|^2$ where $x^*$ is the true coordinates of the source and $\tilde x^*(\omega_i)$ is our coordinate estimate in the $i$-th experiment. In the subsequence, we treat the average squared error as an approxiate of the mean squared error (MSE), i.e.,
	\begin{equation}\label{eqn-MSE-appro}
	\mathbb E\left[\left\| {\tilde x^* - x^*} \right\|^2\right]
	\approx
	\frac{1}{N}\sum_{i=1}^{N}{\left\| {\tilde x^*(\omega_i) - x^*} \right\|^2}.
	\end{equation}
	The performance comparison among our algorithm and the other methods is depicted in Fig.~\ref{setup_1}, where we take X-axis as $10\log(1/\sigma^2)$ and Y-axis as the root mean squared error (RMSE). We see from the figure that when the intensity of the noises is low, our algorithm performs comparable with the CFS and SUM-MPR algorithms, which are superior to the ULS and SDP-I methods. With the increase of noise intensity, our algorithm has the lowest RMSE. 
	\begin{figure} [!htb]
		\centering
		\includegraphics[width=0.48\textwidth]{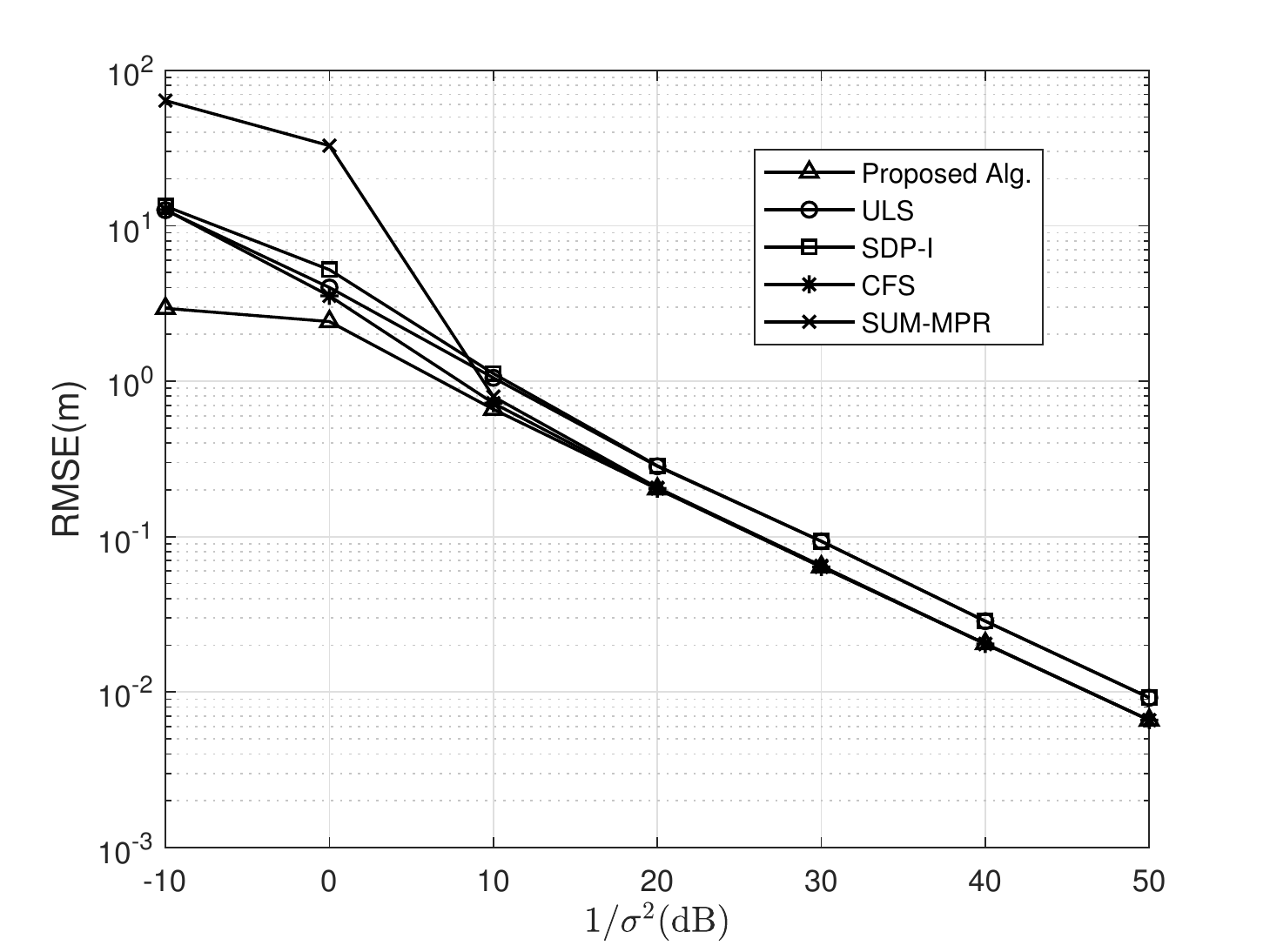}
		\caption{Performance comparison in Example 6.}
		\label{setup_1}
	\end{figure}
	
	\emph{Example 7.} In this example, we leave the other settings unchanged but move four sensors far away from the origin by subtracting $[100~100]^\top$ from the coordinates in Example 1. The results are plotted in Fig.~\ref{setup_2}. In this setup, our estimator outperforms the other estimators by several orders of magnitude. This may be because the distances from the other four sensors to the origin are much larger than those between one another, making the matrix $A$ ill-conditioned. As such, the ULS solution $(A^\top A)^{-1}A^\top b$ will be far away from the feasible set $\{y \mid y^\top D y=0\}$ and has a poor performance. For the SDP-I and CFS algorithms, noticing that their performances in both examples are nearly the same as that of the ULS solution.

	\begin{figure} [!htb]
		\centering
		\includegraphics[width=0.48\textwidth]{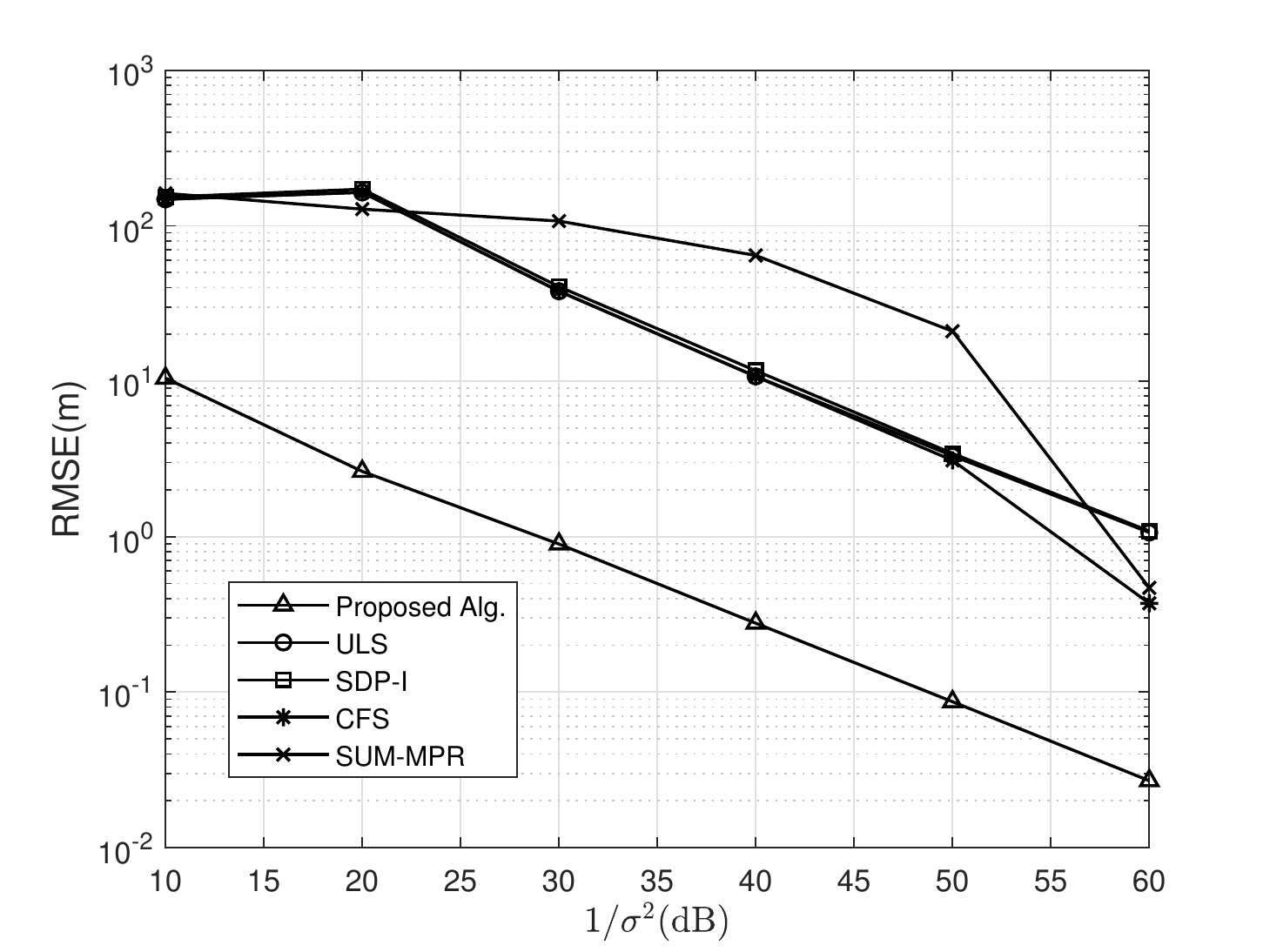}
		\caption{Performance comparison in Example 7.}
		\label{setup_2}
	\end{figure}

}
}

\section{Conclusions}\label{section:conclusions}
In this work, we were concerned about the problem of radiating source localization from range-difference measurements. We placed the attention to a spherical least squares approach that squares the range-difference measurements.
By utilizing this model, one can formulate a CLS range-difference based localization problem, which is a nonconvex optimization problem.
Our first result suggested that the resulting localization problem have bounded global solutions under some rank condition. A necessary and sufficient condition for a global CLS solution was derived by means of the Lagrange multiplier technique.
The uniqueness of a global solution can be equivalently checked using a stricter second-order condition. Consequently, by examining these two conditions, we have complete knowledge of
the multiplicity nature of CLS solutions. Finally, we studied the structural properties of global CLS solutions for some special cases. Our results contribute to
finding out the locations of the global solutions in a convenient manners. Numerical algorithms for computing the CLS solutions by means of our
research findings worth future research. In addition, analysis on asymptotic properties of the statistical estimator generated from the CLS solution will be an interesting problem.

\section*{Appendix A. Proof. of Lemma~\ref{lemma:solution-existence}}\label{appendix:A}

For any vector $y$ such that $y^\top A^\top A y=0$, where $A$ is given in~\eqref{eqn:def-A}, it holds that
$f(y)$ is constant. In addition, since $ A^\top A \geq 0$, for any unbounded sequence $(y_k)_{k\in \mathbb N}$, $\lim \inf _{k\to \infty} f(y_k)$ is unbounded or constant, therefore the CLS problem has at least a global solution in $\mathbb R^{n+1}$.

To show the rest, consider nonzero vectors in a form of $y=[\|x\|,x^{\top}]^\top$ for some $x$ and let the true position of the source be $x^*$. When $y^\top A^\top A y>0$
for all nonzero $y$, for any unbounded such sequence $(y_k)_{k\in \mathbb N}$, $\lim \inf _{k\to \infty} f(y_k)$ is unbounded, which shows the boundedness of the solution set. To evaluate whether $y^\top A^\top A y>0$ is zero or not
for all nonzero $y$, we only need to consider unit vectors, i.e., vectors with $\|x\|=1$. As such, in the rest, we let $\|x\|=1$ in $y$ without loss of generality.
Let $$
\Psi:=
\begin{bmatrix}
\|a_1-x^*\|-\|x^*\|\\
\vdots\\
\|a_m-x^*\|-\|x^*\|
\end{bmatrix}, ~~
\Sigma:=
\begin{bmatrix}
a_1^\top x\\
\vdots\\
a_m^\top x
\end{bmatrix},
$$
and $r:=[r_1,\ldots,r_m]^\top$.
Notice that $Ay=\Psi+\Sigma x+r$. Therefore,
if $r\in \mathcal O:=
\{
-\Sigma x-\Psi:\|x\|=1
\}$, there exists some nonzero $y$ satisfying 
$y^\top A^\top A y=0$. The Lebesgue measure of 
$\Omega$ is zero,  
the solution set is bounded almost surely for any set of $\{d_i,i=1,\ldots,m\}$.

\section*{Appendix B. Proofs of Theorems~\ref{thm:iff-condition} and~\ref{thm2} }
\label{section:appendix_A}

\subsection*{B.1.~Proof of Theorem~\ref{thm:iff-condition}}

We define the Lagrangian function
\begin{equation}\label{eqn:Lagrangian-function}
\mathcal L(y,\lambda)=f(y)+\lambda g(y).
\end{equation}
Notice that $\triangledown_{y} \mathcal L(y,\lambda)$
is well defined.
Next we divide the proof into two steps.

\noindent{\bf Necessity.~} Suppose $y^*$ is a global optimal solution of the
optimization problem~\eqref{eqn:opt-problem-LS-trans}. We consider two cases.

When $y^*\not =0$, the conditions $g(y^*) =0$ and $[y^*]_1>0$ are readily met, and linear independence condition qualification is also satisfied. 
The KKT optimality condition
ensures that there exists a multiplier $\lambda^*$
such that $\triangledown_{ y} \mathcal L(y^*,\lambda^*)=0$, the condition~\eqref{eqn:condition-1}
follows. It further leads to the following result:
for any $y$ with $[y]_1\geq 0$,
\begin{equation}\label{eqn:Lagrangian-function}
\mathcal L(y,\lambda^*)=\mathcal L(y^*,\lambda^*)+
(y-y^*)^\top (A^\top A+\lambda^* D)(y-y^*).
\end{equation}
Since $\mathcal L(y,\lambda^*)=f(y)$ whenever $g(y)=0$, we have that
\begin{equation}\label{eqn:optimal-solution-implication}
v^\top(A^\top A+\lambda^* D) v\geq 0
\end{equation}
holds for $v\in \mathcal T:=\{v: g(y^*+v)=0,
 [v]_1\geq -[y^*]_1\}$. We will use this fact
to show~\eqref{eqn:indefinite-condition}.
To this end, denote $\mathcal S:=\{v: v^\top D v\leq 0\}$ and
partition $\mathcal S$ into  a few disjoint subsets, where
\begin{equation}\label{def:s1-set}
\mathcal S_1=\{v: v^\top D y^*>0, v^\top  Dv<0 \},
\end{equation}
\begin{equation}\label{def:s2-set}
\mathcal S_2=\{v: v^\top D y^*<0, v^\top Dv<0 \},
\end{equation}
\begin{equation}\label{def:s1-set3}
\mathcal S_3=\{v: v^\top D y^*=0, v^\top Dv= 0\}
\end{equation}
and
$S_4=\mathcal S\setminus(\mathcal S_1 \cup \mathcal S_2\cup \mathcal S_3).$

For any $v\in\mathcal S_1$, there always exists a constant $\psi>0$ such that
$g(y^*+\psi v)=g(y^*)+2\psi v^\top D y^*+{\psi}^{2}v^\top D v=g(y^*)=0$,
Next we will show $[\psi v]_1>-[y^*]_1$. Suppose that
$[y^*]_1+[\psi v]_1\leq 0$ were true. Then $([y^*]_1+[\psi v]_1)[y^*]_1\leq 0$, by which we further have
\begin{align*}
& v^\top D y^*\\
=&\frac{1}{\psi}(y^*+\psi v)^\top D y^*\\
=&\frac{1}{\psi}([y^*]_1+[\psi v]_1)[y^*]_1-
\frac{1}{\psi}[y^*+\psi v]_{2:n+1}^\top[y^*]_{2:n+1}\\
\leq& \frac{1}{\psi}([y^*]_1+[\psi v]_1)[y^*]_1+\frac{1}{\psi}
\|[y^*+\psi v]_{2:n+1}\|\|[y^*]_{2:n+1}\|\\
=&0.
\end{align*}
The above inequality contradicts $v\in\mathcal S_1$. Therefore, it is clear that
$\psi v\in\mathcal T$. Then by~\eqref{eqn:optimal-solution-implication}, we conclude that
$$
v^\top(A^\top A+\lambda^* D) v\geq 0
$$
for any $v\in\mathcal S_1$.

For any $v\in\mathcal S_2$, we obtain that $-v\in\mathcal S_1$. Therefore,
$v^\top(A^\top A+\lambda^* D) v\geq 0$ holds
for any $v\in\mathcal S_2$.
Similarly, for any $v\in\mathcal S_3$, we have that $g(y^*+ \phi v)=0$ for any
$\phi\in\mathbb R$ and $[\phi v]_1+[y^*]_1\geq 0$ when $\phi$ is sufficiently small.
It implies that $\phi v \in\mathcal T$. Thus, we have $v^\top(A^\top A+\lambda^* D) v\geq 0$ for any $v\in\mathcal S_3$.

Finally, since each element of $\mathcal S_4$ is a limit point of $\mathcal S_1$, $\mathcal S_2$ or $\mathcal S_3$, the continuity of $v^\top (A^\top A+\lambda^* D)v$ in $v$
implies that $v^\top(A^\top A+\lambda^* D) v\geq 0$ also holds for $v\in\mathcal S_4$.
So far we have shown~\eqref{eqn:indefinite-condition} for the case of $y^*\not =0$.

When $y^*=0$, we will prove the result by contradiction. Suppose that there exists at least a $\tilde   y\not =0$ such that $g(\tilde y)=0$ and $[\tilde y]_1>0$, and there corresponds a $\lambda^*$ satisfying~\eqref{eqn:iff-condition}.
Then by~\eqref{eqn:Lagrangian-function} we have
\begin{align*}
f(0)=& \mathcal L(0,\lambda^*) \\
=&\mathcal L(\tilde y,\lambda^*)+
\tilde y^\top (A^\top A +\lambda^* D) \tilde y \\
=&
f(\tilde y)+
\tilde y^\top (A^\top A +\lambda^* D) \tilde y.
\end{align*}
Since $\tilde y^\top  D \tilde y=0$ and by
Assumption~\ref{lemma:sufficient-global-minimizer},
$$
\tilde y^\top (A^\top A +\lambda D) \tilde y>0.
$$
It thus follows that $f(0)>f(\tilde y)$, which contradicts the hypothesis.

\noindent{\bf Sufficiency.~}
Suppose that $y^*\in \mathbb R^{n+1}$ and $\lambda^*\in\mathbb R$ is a vector and a
multiplier, respectively, which satisfy $(i)$.
Then we have $\triangledown_{y} \mathcal L(y^*,\lambda^*)=0$, which further
yields the result of~\eqref{eqn:Lagrangian-function}.
Let $y\not =y^*$ be any given vector satisfying
$g(y)=0$ and $[y]_1\geq 0$. Then $y$ and $y^*$ have the following
geometric relation
\begin{align}\label{eqn:gy-0-condition-1}
\triangledown g(y^*)(y-y^*)&=2{y^*}^\top D  (y-y^*)\\\notag
& = 2{y^*}^\top D  y\\\notag
&= 2[y^*]_1y_1-2([y^*]_{2:n})^\top [y]_{2:n}\\\notag
&\geq 2[y^*]_1y_1-2\|[y^*]_{2:n}\| \|[y]_{2:n}\|=0.
\end{align}
This also implies
\begin{equation}\label{eqn:gy-0-condition-2}
(y-y^*)^\top\triangledown^2 g(y^*)(y-y^*)=-2{y^*}^\top Dy\leq 0.
\end{equation}
In other words, by~\eqref{eqn:gy-0-condition-2},
we can find a vector $v=y-y^*$ satisfying
$v^\top\triangledown^2 g(y^*)v\leq 0$,
which together with~\eqref{eqn:Lagrangian-function}
implies that
\begin{align*}
\mathcal L(y,\lambda^*)&=\mathcal L(y^*,\lambda^*)+
v^\top (A^\top A+\lambda^* D)v\\
&\geq \mathcal L(y^*,\lambda^*),
\end{align*}
where the inequality follows from~\eqref{eqn:indefinite-condition}.
Observing that $g(y)=g(y^*)=0$, then the above inequality yields that
$$
f(y^*)\leq f(y),
$$
which shows that $y^*$ is a global optimal solution of  problem~\eqref{eqn:opt-problem-LS-trans}.

Next we will show that $(ii)$ is a sufficient condition for the optimality of $y$.
Suppose that $y^*=0$ is not an optimal solution to problem~\eqref{eqn:opt-problem-LS-trans}. By Lemma~\ref{lemma:solution-existence},
we can find at least a vector $\tilde y\not =0$ that is a global minimizer of~\eqref{eqn:opt-problem-LS}. By the argument used for showing the {necessity},
we conclude that~the conditions in~\eqref{eqn:iff-condition} hold for $\tilde y$, which contradicts $(ii)$.
\hfill$\blacksquare$

\subsection*{B.2.~Proof of Theorem~\ref{thm2}}
The proof resembles that of Theorem~\ref{thm:iff-condition}. Here we only provide with a sketch.

\noindent{\bf Sufficiency.~} We first suppose that $y^*\in \mathbb R^{n+1}$ and $\lambda^*\in\mathbb R$ is a vector and a
multiplier, respectively, that satisfy $(i)$.
Then we have $\triangledown_y \mathcal L(y^*,\lambda^*)=0$, further
yielding that, for any $y$ with $[y]_1\geq 0$, the equality~\eqref{eqn:Lagrangian-function}
holds.
For any given vector $y\not =y^*$ satisfying
$g(y)=0$ and $[y]_1\geq 0$, by~\eqref{eqn:gy-0-condition-2}, we have $(y-y^*)^\top\triangledown^2 g(y^*)(y-y^*)\leq 0$.
When $(y-y^*)^\top\triangledown^2 g(y^*)(y-y^*)< 0$,
we have
\begin{align*}
(y-y^*)^\top  Dy^*&=y^\top  Dy^*\\
&>[y^*]_1y_1-\|[y^*]_{2:n}\| \|[y]_{2:n}\|=0.
\end{align*}
By~\eqref{eqn:Lagrangian-function} and~\eqref{eqn:indefinite-condition-unique},
\begin{align*}
f(y)= &\mathcal L(y,\lambda^*) \\
=&\mathcal L(y^*,\lambda^*)+
(y-y^*)^\top (A^\top A+\lambda^* D)(y-y^*) \\
>& \mathcal L(y^*,\lambda^*)=f(y^*);
\end{align*}
and when $(y-y^*)^\top\triangledown^2 g(y^*)(y-y^*)= 0$, due to Assumption~\ref{lemma:sufficient-global-minimizer},
\begin{align*}
f(y)&=\mathcal L(y,\lambda^*)\\
&=\mathcal L(y^*,\lambda^*)+
(y-y^*)^\top (A^\top A+\lambda^* D)(y-y^*)\\
&=\mathcal L(y^*,\lambda^*)+
(y-y^*)^\top A^\top A(y-y^*)\\
&>\mathcal L(y^*,\lambda^*)=f(y^*).
\end{align*}
The above altogether imply that $y^*$ is  the unique global optimal solution
of~\eqref{eqn:opt-problem-LS}. For $(ii)$, notice that $y=0$ is a minimizer in this situation and that
\eqref{eqn:iff-condition} is a necessary condition for
a $y\not =0$ being an solution to
problem~\eqref{eqn:opt-problem-LS-trans}, both by Theorem~\ref{thm:iff-condition}.
They altogether show that the origin is the unique solution.

\noindent{\bf Necessity.~} First consider $y^*\not =0$.
Since $y^*$ is the unique optimal solution,
similar to~\eqref{eqn:optimal-solution-implication},
we have that $v^\top(A^\top A+\lambda^* D) v> 0$
holds for $v\in \mathcal T\setminus\{0\}$.
Since $\{v: v\top D v<0, v^\top D y^*\neq 0\}=\mathcal S_1 \cup \mathcal S_2$, where
$\mathcal S_1 $ and $\mathcal S_2$ are given in~\eqref{def:s1-set} and~\eqref{def:s2-set}, following the proof of Theorem~\ref{thm:iff-condition},
we eventually obtain $(i)$. For $y^*=0$, the proof directly follows from that of
Theorem~\ref{thm:iff-condition}.

The above derivations altogether conclude the claim.\hfill$\blacksquare$

\section*{Appendix C. Proofs of lemmas in Section~\ref{section:solution_algorithm}}
\label{section:Appendix_B}

This appendix contains proofs of some lemmas in Section~\ref{section:solution_algorithm}.

\subsection*{C.1.~Proof of Lemma~\ref{lemma:distinct-solutions}}

When $h(\lambda)<0$ for $\lambda\in\mathcal I_1$,
it has been shown in~\cite{More93OMS}
that $g(z^-)>0$. Since $g(y(\lambda))=h(\lambda)$, by continuity we have
\begin{equation}\label{eqn:g_y_leq_0}
g(y_*)\leq 0.
\end{equation}
We consider the following two cases, respectively.

Suppose that $g(y_*)< 0$. It is straightforward to see that~\eqref{eqn:quadratic-eqn}
has two different roots: one is positive and the other is negative.

Suppose that $g(y_*)= 0$. First we shall show that $y_*^\top D z^-\not =0$.
Computation suggests that
\begin{align*}
|y_*^\top D z^-|= &|[y_*]_1[z^-]_1-[y_*]_{2:n+1}^\top [z^-]_{2:n+1}|\\
\geq &|[y_*]_1[z^-]_1|-|[y_*]_{2:n+1}^\top [z^-]_{2:n+1}|\\
\geq &|[y_*]_1[z^-]_1|-\|[y_*]_{2:n+1}\|\| [z^-]_{2:n+1}\|
\end{align*}
Observe that $|[y_*]_1|=\|[y_*]_{2:n+1}\|$ due to $g(y_*)= 0$ and
 $|[z^-]_1|>\|[z^-]_{2:n+1}\|$ due to $g(z^-)> 0$. Combining these
facts, we obtain that $|y_*^\top D z^-|>0$, implying
\begin{equation}\label{eqn:non-zero-yDz}
y_*^\top D z^-\not = 0.
\end{equation}
 Since $g(y_*)= 0$ and $g(z^-)>0$, we finally
conclude that~\eqref{eqn:quadratic-eqn} has a zero root and a nonzero root.

We conclude the result by completing the analysis for the above two cases.\hfill$\blacksquare$

\subsection*{C.2.~Proof of Lemma~\ref{lemma:sign-of-y-lambda-l}}

The following technical lemma is functional to the proof .
\begin{lemma}\label{lemma:convex-combination-inequality}
Let $y, \bar y\in\mathbb R^{n+1}$ be two vectors satisfying
$g(y)=g(\bar y)=0$.
Then the following statement are true:
\begin{enumerate}
\item [(i).] If $[y]_1[\bar y]_1\geq 0$, it holds that $g(\psi y+(1-\psi)\bar y)\geq 0$ for any $\psi\in[0,1]$.

\item [(ii).] If $[y]_1[\bar y]_1\leq 0$,
it holds that $g(\psi y+(1-\psi)\bar y)\leq 0$ for any $\psi\in[0,1]$.
\end{enumerate}
\end{lemma}
\noindent \textbf{Proof.~}
First we prove (i).
We begin with an observation that
\begin{align*}
& g(\psi y+(1-\psi)\bar y) \\
=&\psi^2 g(y)+(1-\psi)^2g(\bar y)+2\psi(1-\psi)y^\top D \bar y\\
=& 2\psi(1-\psi)[y]_1[\bar y]_1-
2\psi(1-\psi)[y]_{2:n+1}^\top [\bar y]_{2:n+1}\\
\geq& 2\psi(1-\psi)[y]_1[\bar y]_1
-2\psi(1-\psi)\|[y]_{2:n+1}\|\|[\bar y]_{2:n+1}\|\\
=&0,
\end{align*}
where the last inequality follows from the Cauchy-Schwarz inequality~\cite{Rudin1964realanalysis} and the last equality
holds due to the form of $D$ given in~\eqref{eqn:def-D} and
$g(y)=g(\bar y)=0$.

To show (ii), similarity, we have
\begin{align*}
& g(\psi y+(1-\psi)\bar y)\\
=& 2\psi(1-\psi)[y]_1[\bar y]_1-
2\psi(1-\psi)[y]_{2:n+1}^\top [\bar y]_{2:n+1}\\
\leq & 2\psi(1-\psi)[y]_1[\bar y]_1
+2\psi(1-\psi)\|[y]_{2:n+1}\|\|[\bar y]_{2:n+1}\|\\
=&0,
\end{align*}
which completes the proof.
\hfill$\blacksquare$

\noindent \textbf{Proof of Lemma~\ref{lemma:sign-of-y-lambda-l}.~}
It is clear that neither $y_*+\alpha_1^-z^-$ nor $y_*+\alpha_2^-z^-$
is $0$, since otherwise~\eqref{eqn:condition-1} is violated given that
$b\not=0$. Notice that $y_*$ can be expressed as
a linear combination of $y_*+\alpha_1^-z^-$ and $y_*+\alpha_2^-z^-$
as follows:
\begin{equation}\label{eqn:convex_combination}
y_*=\frac{\alpha_1^-}{\alpha_1^--\alpha_2^-}
(y_*+\alpha_2^-z^-)+\frac{-\alpha_2^-}{\alpha_1^--\alpha_2^-}
(y_*+\alpha_1^-z^-).
\end{equation}
In~\eqref{eqn:convex_combination}, $\alpha_1^-$ and $\alpha_2^-$ cannot happen to be
both positive or both negative. Therefore, $\frac{-\alpha_2^-}{\alpha_1^--\alpha_2^-},
\frac{\alpha_1^-}{\alpha_1^--\alpha_2^-}\in[0,1]$ and the above combination is a
convex one.

In virtue of~\eqref{eqn:g_y_leq_0}, we will proceed by considering
the case in which the equality sign of~\eqref{eqn:g_y_leq_0} holds and the
case in which the strict inequality sign of~\eqref{eqn:g_y_leq_0} holds, respectively.
Suppose that $g(y_*)<0$. Then, the form of~\eqref{eqn:quadratic-eqn} suggest that
$\alpha_1^->0$ and $\alpha_2^-<0$. If $[y_*+\alpha_1^-z^-]_1[y_*+\alpha_2^-z^-]_1 \geq 0$, then by Lemma~\ref{lemma:convex-combination-inequality}
$g(y_*)\geq 0$, which contradicts the hypothesis.
Suppose that $g(y_*)=0$. By~\eqref{eqn:non-zero-yDz}, we consider two cases, i.e., $y_*^\top D z^->0$ and $y_*^\top D z^-<0$, respectively. First, assume that
$y_*^\top D z^->0$. Then $\alpha_1^-=0$ and $\alpha_2^-<0$.
If $[y_*]_1[y_*+\alpha_2^-z^-]_1$
were larger than $0$, it suggests from computation that
\begin{align*}
y_*^\top D z^-=&\frac{\alpha_2^- y_*^\top D z^-}{\alpha_2^-}\\
=&\frac{y_*^\top D (y_*+\alpha_2^- z^-)}{\alpha_2^-}\\
=&\frac{1}{\alpha_2^-}([y_*]_1 [y_*+\alpha_2^- z^-]_1 \\
&-[y_*]_{2:n+1}^\top [y_*+\alpha_2^- z^-]_{2:n+1}) \\
\leq & \frac{1}{\alpha_2^-}([y_*]_1 [y_*+\alpha_2^- z^-]_1 \\
&-\|[y_*]_{2:n+1}\| \|[y_*+\alpha_2^- z^-]_{2:n+1}\|)\\
=&0,
\end{align*}
where the last equality follows since $g(y_*)=g(y_*+\alpha_2^- z^-)=0$.
It contradicts the hypothesis. On the other hand, when we assume
that $y_*^\top D z^-<0$, we reach a similar contradiction, which concludes
the result. \hfill$\blacksquare$

\subsection*{C.3.~Proof of Lemma~\ref{lemma:sign-of-y-lambda-u}}

When $h(\lambda)>0$ for $\lambda\in\mathcal I_1$,
it has been shown in~\cite{More93OMS}
that $g(z^+)<0$. In addition, by continuity, we have $g(y_*)\geq 0.$

The first claim can be proved by contradiction. The procedure is similar to
the proof of Lemma~\ref{lemma:sign-of-y-lambda-l} and is omitted.

Next we prove the second claim. As $\alpha_1^+$ and $\alpha_2^+$  are the roots of~\eqref{eqn:quadratic-eqn-2}.
Therefore,
$\alpha_1^+$ and $\alpha_2^+$ are not
both positive or both negative since $g(y_*)/g(z^+)\leq 0$.
Then $y_*$ is a convex combination
of $y_*+\alpha_1^+z^+$ and $y_*+\alpha_2^+z^+$. It further implies that
$[y_*]_1$, $[y_*+\alpha_1^+z^+]_1$ and $[y_*+\alpha_2^+z^+]_1$
have the same sign. In addition, $y(\lambda)$ is a rational function for
$\lambda\in\mathcal I_1$~\cite{More93OMS} and $\mathcal I_1$ is a connected set.
We then obtain that
the image of $\mathcal I_1$ under the function $y$, denoted by $y[\mathcal I_1]$, is connected. If there were a $\lambda'\in \mathcal I_1$
such that $[y(\lambda')]_1$ and $[y_*]_1$ have distinct signs, we could find
another $\lambda^{''}\in\mathcal I_1$ satisfying $[y(\lambda^{''})]_1=0$.
Notice  that $g(y(\lambda^{''}))=-\|[y(\lambda^{''})]_{2:n+1}\|^2\leq 0$, which contradicts
$g(y(\lambda^{''})):=h(\lambda^{''})>0$ for all $\lambda\in\mathcal I_1$
and completes the proof. \hfill$\blacksquare$


\begin{thebibliography}{10}
\providecommand{\url}[1]{#1}
\csname url@samestyle\endcsname
\providecommand{\newblock}{\relax}
\providecommand{\bibinfo}[2]{#2}
\providecommand{\BIBentrySTDinterwordspacing}{\spaceskip=0pt\relax}
\providecommand{\BIBentryALTinterwordstretchfactor}{4}
\providecommand{\BIBentryALTinterwordspacing}{\spaceskip=\fontdimen2\font plus
\BIBentryALTinterwordstretchfactor\fontdimen3\font minus
  \fontdimen4\font\relax}
\providecommand{\BIBforeignlanguage}[2]{{%
\expandafter\ifx\csname l@#1\endcsname\relax
\typeout{** WARNING: IEEEtran.bst: No hyphenation pattern has been}%
\typeout{** loaded for the language `#1'. Using the pattern for}%
\typeout{** the default language instead.}%
\else
\language=\csname l@#1\endcsname
\fi
#2}}
\providecommand{\BIBdecl}{\relax}
\BIBdecl


\bibitem{tiemann2017scalable}
J.~Tiemann and C.~Wietfeld, ``Scalable and precise multi-UAV indoor navigation using TDOA-based UWB localization,''.\hskip 1em plus 0.5em minus 0.4em\relax \emph{ In Proceedings of the international conference on indoor positioning and indoor navigation (IPIN)}. IEEE, 2017, pp, 1--7.

\bibitem{bottigliero2021low}
S.~Bottigliero, D.~Milanesio, M.~Saccani and R.~Maggiora, ``A Low-Cost Indoor Real-Time Locating System Based on TDOA Estimation of UWB Pulse Sequences", \emph{IEEE Transactions on Instrumentation and Measurement}, vol.~70, pp. 1--11, 2021.

\bibitem{gentile2012geolocation}
C.~Gentile, N.~Alsindi, R.~Raulefs, and C.~Teolis, \emph{Geolocation
  Techniques: Principles and Applications}.\hskip 1em plus 0.5em minus
  0.4em\relax Springer Science \& Business Media, 2012.

\bibitem{mao2007wireless}
G.~Mao, B.~Fidan, and B.~D. Anderson, ``Wireless sensor network localization
  techniques,'' \emph{Computer networks}, vol.~51, no.~10, pp. 2529--2553,
  2007.
  
\bibitem{martalo2021improved}
M.~Martalo, S.~Perri, G.~Verdano and F.~De Mola, F.~Monica and G.~Ferrari, ``Improved UWB TDoA-based Positioning using a Single Hotspot for Industrial IoT Applications",
\emph{IEEE Transactions on Industrial Informatics}, 2021.

\bibitem{raza2019dataset}
U.~Raza, A.~Khan, R.~Kou, T.~Farnham, T.~Premalal, A.~Stanoev and W.~Thompson,
``Dataset: Indoor Localization with Narrow-band, Ultra-Wideband, and Motion Capture Systems",
\emph{in Proceedings of the 2nd Workshop on Data Acquisition to Analysis}, pp.~34--36, 2019.

\bibitem{sidorenko2020error}
J.~Sidorenko, V.~Schatz, N.~Scherer-Negenborn, M.~Arens and U.~Hugentobler,``Error corrections for ultrawideband ranging", \emph{IEEE Transactions on Instrumentation and Measurement}, vol.~69, no.~11, pp.~9037--9047, 2020.


\bibitem{win2011network}
M.~Z. Win, A.~Conti, S.~Mazuelas, Y.~Shen, W.~M. Gifford, D.~Dardari, and
  M.~Chiani, ``Network localization and navigation via cooperation,''
  \emph{IEEE Communications Magazine}, vol.~49, no.~5, 2011.

\bibitem{vermesan2011internet}
O.~Vermesan, P.~Friess, P.~Guillemin, S.~Gusmeroli, H.~Sundmaeker, A.~Bassi,
  I.~S. Jubert, M.~Mazura, M.~Harrison, M.~Eisenhauer \emph{et~al.}, ``Internet
  of things strategic research roadmap,'' \emph{Internet of Things-Global
  Technological and Societal Trends}, vol.~1, no. 2011, pp. 9--52, 2011.

\bibitem{papadimitratos2009vehicular}
P.~Papadimitratos, A.~De~La~Fortelle, K.~Evenssen, R.~Brignolo, and S.~Cosenza,
  ``Vehicular communication systems: Enabling technologies, applications, and
  future outlook on intelligent transportation,'' \emph{IEEE communications
  magazine}, vol.~47, no.~11, 2009.

\bibitem{constandache2010did}
I.~Constandache, X.~Bao, M.~Azizyan, and R.~R. Choudhury, ``Did you see bob?:
  human localization using mobile phones,'' in \emph{Proceedings of the
  sixteenth annual international conference on Mobile computing and
  networking}.\hskip 1em plus 0.5em minus 0.4em\relax ACM, 2010, pp. 149--160.

\bibitem{langendoen2003distributed}
K.~Langendoen and N.~Reijers, ``Distributed localization in wireless sensor
  networks: a quantitative comparison,'' \emph{Computer networks}, vol.~43,
  no.~4, pp. 499--518, 2003.

\bibitem{priyantha2003anchor}
N.~B. Priyantha, H.~Balakrishnan, E.~Demaine, and S.~Teller, ``Anchor-free
  distributed localization in sensor networks,'' in \emph{Proceedings of the
  1st international conference on Embedded networked sensor systems}.\hskip 1em
  plus 0.5em minus 0.4em\relax ACM, 2003, pp. 340--341.

\bibitem{khan2009distributed}
U.~A. Khan, S.~Kar, and J.~M. Moura, ``Distributed sensor localization in
  random environments using minimal number of anchor nodes,'' \emph{IEEE
  Transactions on Signal Processing}, vol.~57, no.~5, pp. 2000--2016, 2009.

\bibitem{youssef2005horus}
M.~Youssef and A.~Agrawala, ``The horus wlan location determination system,''
  in \emph{Proceedings of the 3rd international conference on Mobile systems,
  applications, and services}.\hskip 1em plus 0.5em minus 0.4em\relax ACM,
  2005, pp. 205--218.

\bibitem{chintalapudi2010indoor}
K.~Chintalapudi, A.~Padmanabha~Iyer, and V.~N. Padmanabhan, ``Indoor
  localization without the pain,'' in \emph{Proceedings of the sixteenth annual
  international conference on Mobile computing and networking}.\hskip 1em plus
  0.5em minus 0.4em\relax ACM, 2010, pp. 173--184.

\bibitem{rai2012zee}
A.~Rai, K.~K. Chintalapudi, V.~N. Padmanabhan, and R.~Sen, ``Zee: Zero-effort
  crowdsourcing for indoor localization,'' in \emph{Proceedings of the 18th
  annual international conference on Mobile computing and networking}.\hskip
  1em plus 0.5em minus 0.4em\relax ACM, 2012, pp. 293--304.

\bibitem{dissanayake2001solution}
M.~G. Dissanayake, P.~Newman, S.~Clark, H.~F. Durrant-Whyte, and M.~Csorba, ``A
  solution to the simultaneous localization and map building (slam) problem,''
  \emph{IEEE Transactions on robotics and automation}, vol.~17, no.~3, pp.
  229--241, 2001.

\bibitem{durrant2006simultaneous}
H.~Durrant-Whyte and T.~Bailey, ``Simultaneous localization and mapping: part
  i,'' \emph{IEEE robotics \& automation magazine}, vol.~13, no.~2, pp.
  99--110, 2006.

\bibitem{chandrasekhar2006localization}
V.~Chandrasekhar, W.~K. Seah, Y.~S. Choo, and H.~V. Ee, ``Localization in
  underwater sensor networks: survey and challenges,'' in \emph{Proceedings of
  the 1st ACM international workshop on Underwater networks}.\hskip 1em plus
  0.5em minus 0.4em\relax ACM, 2006, pp. 33--40.

\bibitem{dil2006range}
B.~Dil, S.~Dulman, and P.~Havinga, ``Range-based localization in mobile sensor
  networks,'' in \emph{European Workshop on Wireless Sensor Networks}.\hskip
  1em plus 0.5em minus 0.4em\relax Springer, 2006, pp. 164--179.

\bibitem{seco2009survey}
F.~Seco, A.~R. Jim{\'e}nez, C.~Prieto, J.~Roa, and K.~Koutsou, ``A survey of
  mathematical methods for indoor localization,'' in \emph{2009 IEEE
  International Symposium on Intelligent Signal Processing}.\hskip 1em plus
  0.5em minus 0.4em\relax IEEE, 2009, pp. 9--14.

\bibitem{guvenc2009survey}
I.~Guvenc and C.-C. Chong, ``A survey on toa based wireless localization and
  nlos mitigation techniques,'' \emph{IEEE Communications Surveys \&
  Tutorials}, vol.~11, no.~3, 2009.

\bibitem{musicki2010mobile}
D.~Musicki, R.~Kaune, and W.~Koch, ``Mobile emitter geolocation and tracking
  using tdoa and fdoa measurements,'' \emph{IEEE Transactions on Signal
  Processing}, vol.~58, no.~3, pp. 1863--1874, 2010.

\bibitem{salameh2010cooperative}
H.~A.~B. Salameh, M.~Krunz, and O.~Younis, ``Cooperative adaptive spectrum
  sharing in cognitive radio networks,'' \emph{IEEE/ACM Transactions On
  Networking}, vol.~18, no.~4, pp. 1181--1194, 2010.

\bibitem{karbasi2012robust}
A.~Karbasi and S.~Oh, ``Robust localization from incomplete local
  information,'' \emph{IEEE/ACM Transactions on Networking}, vol.~21, no.~4,
  pp. 1131--1144, 2012.

\bibitem{liu2007survey}
H.~Liu, H.~Darabi, P.~Banerjee, and J.~Liu, ``Survey of wireless indoor
  positioning techniques and systems,'' \emph{IEEE Transactions on Systems,
  Man, and Cybernetics, Part C (Applications and Reviews)}, vol.~37, no.~6, pp.
  1067--1080, 2007.

\bibitem{bruck2009localization}
J.~Bruck, J.~Gao, and A.~A. Jiang, ``Localization and routing in sensor
  networks by local angle information,'' \emph{ACM Transactions on Sensor
  Networks (TOSN)}, vol.~5, no.~1, p.~7, 2009.

\bibitem{kaplan2005understanding}
E.~Kaplan and C.~Hegarty, \emph{Understanding GPS: principles and
  applications}.\hskip 1em plus 0.5em minus 0.4em\relax Artech house, 2005.

\bibitem{misra2006global}
P.~Misra and P.~Enge, ``Global positioning system: Signals, measurements and
  performance second edition,'' \emph{Massachusetts: Ganga-Jamuna Press}, 2006.

\bibitem{ash2004sensor}
J.~Ash and L.~Potter, ``Sensor network localization via received signal
  strength measurements with directional antennas,'' in \emph{Proceedings of
  the 2004 Allerton Conference on Communication, Control, and Computing}, 2004,
  pp. 1861--1870.

\bibitem{Stoica2006}
P.~Stoica and J.~Li, ``Source localization from range-difference
  measurements,'' \emph{IEEE Signal Processing Magazine}, vol.~23, no.~6, pp.
  63--66, 2006.

\bibitem{More93OMS}
J.~J. Mor$\rm\acute{e}$, ``Generalizations of the trust region problem,''
  \emph{Optimization methods and Software}, vol.~2, no. 3-4, pp. 189--209,
  1993.

\bibitem{Beck2008}
A.~Beck, P.~Stoica, and J.~Li, ``Exact and approximate solutions of source
  localization problems,'' \emph{IEEE Transactions on Signal Processing},
  vol.~56, no.~5, pp. 1770--1778, 2008.

\bibitem{ljung1983theory}
L.~Ljung and T.~S{\"o}derstr{\"o}m, \emph{Theory and practice of recursive
  identification}.\hskip 1em plus 0.5em minus 0.4em\relax MIT press, 1983.

\bibitem{leonard2012directed}
J.~J. Leonard and H.~F. Durrant-Whyte, \emph{Directed sonar sensing for mobile
  robot navigation}.\hskip 1em plus 0.5em minus 0.4em\relax Springer Science \&
  Business Media, 2012, vol. 175.

\bibitem{sundar2018tdoa}
H.~Sundar, T.~V. Sreenivas, and C.~S. Seelamantula, ``Tdoa-based multiple
  acoustic source localization without association ambiguity,'' \emph{IEEE/ACM
  Transactions on Audio, Speech, and Language Processing}, vol.~26, no.~11, pp.
  1976--1990, 2018.

\bibitem{le2019uncovering}
T.-K. Le and K.~Ho, ``Uncovering source ranges from range differences observed
  by sensors at unknown positions: Fundamental theory,'' \emph{IEEE
  Transactions on Signal Processing}, vol.~67, no.~10, pp. 2665--2678, 2019.

\bibitem{cobos2020frequency}
M.~Cobos, F.~Antonacci, L.~Comanducci, and A.~Sarti, ``Frequency-sliding
  generalized cross-correlation: a sub-band time delay estimation approach,''
  \emph{IEEE/ACM Transactions on Audio, Speech, and Language Processing},
  vol.~28, pp. 1270--1281, 2020.

\bibitem{Foy76TAES}
W.~H. Foy, ``Position-location solutions by taylor-series estimation,''
  \emph{IEEE Transactions on Aerospace and Electronic Systems}, vol.~2, pp.
  187--194, 1976.

\bibitem{Torrieri84TAES}
D.~J. Torrieri, ``Statistical theory of passive location systemss,'' \emph{IEEE
  transactions on Aerospace and Electronic Systems}, vol.~2, pp. 183--198,
  1984.

\bibitem{Smith1987IJOE}
J.~Smith and J.~Abel, ``The spherical interpolation method of source
  localiztion,'' \emph{IEEE Journal Oceanic Engineering}, vol. OE-13, 1987.

\bibitem{smith1987closed}
------, ``Closed-form least-squares source location estimation from
  range-difference measurements,'' \emph{IEEE Transactions on Acoustics,
  Speech, and Signal Processing}, vol.~35, no.~12, pp. 1661--1669, 1987.

\bibitem{friedlander1987passive}
B.~Friedlander, ``A passive localization algorithm and its accuracy analysis,''
  \emph{IEEE Journal of Oceanic engineering}, vol.~12, no.~1, pp. 234--245,
  1987.

\bibitem{Schau1987ITASSP}
H.~C. Schau and A.~Z. Robinson, ``Passive source localization employing
  intersecting spherical surfaces from time-of-arrival differences,''
  \emph{IEEE Transactions on Acoustics, Speech, and Signal Processing},
  vol.~35, no.~8, pp. 1223--1225, 1987.

\bibitem{chan1994simple}
Y.-T. Chan and K.~Ho, ``A simple and efficient estimator for hyperbolic
  location,'' \emph{IEEE Transactions on signal processing}, vol.~42, no.~8,
  pp. 1905--1915, 1994.

\bibitem{huang2001real}
Y.~Huang, J.~Benesty, G.~W. Elko, and R.~M. Mersereati, ``Real-time passive
  source localization: A practical linear-correction least-squares approach,''
  \emph{IEEE transactions on Speech and Audio Processing}, vol.~9, no.~8, pp.
  943--956, 2001.

\bibitem{li2004least}
D.~Li and Y.~H. Hu, ``Least square solutions of energy based acoustic source
  localization problems,'' in \emph{Parallel Processing Workshops, 2004. ICPP
  2004 Workshops. Proceedings. 2004 International Conference on}.\hskip 1em
  plus 0.5em minus 0.4em\relax IEEE, 2004, pp. 443--446.

\bibitem{cheung2006constrained}
K.~W. Cheung, H.-C. So, W.-K. Ma, and Y.-T. Chan, ``A constrained least squares
  approach to mobile positioning: algorithms and optimality,'' \emph{EURASIP
  Journal on Advances in Signal Processing}, vol. 2006, no.~1, p. 020858, 2006.

\bibitem{Gillette08TSPL}
D.~M. Gillette and F.~S. Harvey, ``A linear closed-form algorithm for source
  localization from time-differences of arrival,'' \emph{IEEE Signal Processing
  Letters}, vol.~15, pp. 1--4, 2008.

\bibitem{Giaquinta07Mathematicalanalysis}
M.~Giaquinta and G.~Modica, \emph{Mathematical analysis: linear and metric
  structures and continuity}.\hskip 1em plus 0.5em minus 0.4em\relax Springer
  Science \& Business Media, 2007.

\bibitem{Rudin1964realanalysis}
W.~Rudin, \emph{Principles of mathematical analysis}.\hskip 1em plus 0.5em
  minus 0.4em\relax New York: McGraw-Hill, 1964, vol.~3.

\bibitem{Ljung}  L. Ljung, and T. Glad, ``On global identifiability for arbitrary model parametrizations,'' \emph{Automatica}, vol.~30, pp.~265–276, 1994.


\bibitem{Canclini15IASLP}
 A. Canclini, and P. Bestagini, and F. Antonacci, and M. Compagnoni, and A. Sarti, and S, Tubaro, ``A robust and low-complexity source localization algorithm for asynchronous distributed microphone networks,'' \emph{IEEE/ACM Transactions on Audio, Speech, and Language Processing}, vol.~23, pp.~1563-1575, 2005.


\bibitem{sun18tsp}
Y. Sun, and K. C. Ho, and Q. Wan, ``Solution and analysis of TDOA localization of a near or distant source in closed form,'' \emph{IEEE Transactions on Signal Processing}, vol.~67, pp.~320-335,2018.

\bibitem{huang13twc}
B.~Huang, and L.~Xie, and Z.~Yang, ``TDOA-based source localization with distance-dependent noises,'' \emph{IEEE Transactions on Wireless Communications}, vol.~14, pp.~468-480, 2014.


\bibitem{xu2010reduced}
E.~Xu, and Z.~Ding, and S.~Dasgupta, ``Reduced complexity semidefinite relaxation algorithms for source localization based on time difference of arrival,'' \emph{IEEE Transactions on Mobile Computing},
vol.~10, pp. 1276-1282, 2010.


\bibitem{Ljung94Auto}
L.~Ljung, and G.~Torkel, ``On global identifiability for arbitrary model parametrizations,'' Automatica vol.~30, pp.~265-276,1994.


\bibitem{Rudin76Math_analysis}
W. Rudin, ``Principles of mathematical analysis,'' Vol. 3. New York: McGraw-hill, 1976.

\end{thebibliography}
\end{document}